\documentstyle[11pt,epsfig]{article}
\newcommand{\s}{\sigma }

\newcommand{\xn}{x_{n}}

\newcommand{\kim}{ k_{1}^{\mu}}                                      
\newcommand{\kom}{ k_{0}^{\mu}}                                      
\newcommand{\ki}{ k_{1}}
\newcommand{\yn}{ Y_{n}}                                             
\newcommand{\ym}{ Y_{m}} 
\newcommand{\kn}{ k_{n}}
\newcommand{\Kn}{ K_{n}}
\newcommand{\km}{ k_{m}}
\newcommand{\Km}{ K_{m}}
\newcommand{\kt}{ k_{2}}                                             
\newcommand{\ko}{ k_{0}}

\newcommand{\kin}{ k_{1}^{\nu}}                                      
\newcommand{\kon}{ k_{0}^{\nu}}                                      
\newcommand{\ktm}{ k_{2}^{\nu}}

\newcommand{\lpp}{\mbox {$e^{i\int _{c} \alpha (s) 
k(s) \partial _{z} X(z+s) ds +ik_{0}X(z)}$}}    
\newcommand{\bp }{\mbox {$e^{i\int ~d t~ \int _{c}~ k(s, t )
\partial _{z} X(z (t)+as )~ds+ i\int ~dt~ \ko (t)X(z(t))}$}}
                   
\newcommand{\gvk}{ e^{i\sum _{n }k_{n}Y_{n}(z)}}                      
\newcommand{\gvks}{ e^{i\sum _{n\ge 0 }\int~ dt~ k_{n}(t )\tY _{n}(t )}}

\newcommand{\dds}{\frac{\delta}{\delta \sigma}}

\newcommand{\p}{\partial}                                           
\newcommand{\pp}{\partial ^{2}}                                     
\newcommand{\mup}{\partial _{\mu}}

\newcommand{\eps}{ \epsilon}                                        
\newcommand{\al}{\alpha }                                             
\newcommand{\aln}{\alpha _{n}} 
\newcommand{\tY}{\tilde Y}                                 
\newcommand{\lan}{\langle}
\newcommand{\ran}{\rangle}

\newcommand{\la}{\mbox{$ \lambda $}} 
\newcommand{\be}{\begin{equation}}
\newcommand{\br}{\begin{eqnarray}}
\newcommand{\ee}{\end{equation}} 
\newcommand{\er}{\end{eqnarray}} 
\newcommand{\eln}{\mbox {$ e^{\sum _{n}\lambda _{-n}L_{+n}}$}}

\begin{document} 
\title{
\hfill\parbox{4cm}{\normalsize IMSC/2002/07/16\\
                               hep-th/0207098}\\        
\vspace{2cm}
 Loop Variables in String Theory
\author{B. Sathiapalan\\ {\em Institute of Mathematical Sciences}\\
{\em Taramani}\\{\em Chennai, India 600113}}}                                
\maketitle 

\begin{abstract} 
The loop variable approach is a proposal for a gauge invariant generalization
of the sigma-model renormalization group method of obtaining
equations of motion in string theory. The basic guiding principle
is space-time gauge invariance rather than world-sheet
properties. In essence it is a version of Wilson's exact renormalization
group equation for the world sheet theory. It involves
all the massive modes and is defined with a finite world-sheet cutoff, 
which  allows one to go off the mass-shell. 
On shell the tree amplitudes of string theory are reproduced. The equations
are gauge invariant off shell also. This article is a self-contained
discussion of the loop variable approach as well as its connection 
with the Wilsonian RG.
\end{abstract}
\newpage

\setcounter{equation}{00}

\section{Introduction}
It has been known for some time now that
the renormalization group equations ($\beta$-functions)
for the 2-dimensional action of a string in a non-trivial background
is expected to give the equations of motion for the modes of the string
[\cite{L}-\cite{T}].
This is expected to be 
true both for the closed string modes as well as open string modes.
For massless modes, which were the first to be studied,
this is relatively easy. In certain limits it can be done to all orders \cite{ACNY,FT}.
 For the tachyon also
it has been done in some detail \cite{BSPT}  and in some limits can be done
to all orders \cite{Wi,LW,Sh,KMM}. 

A natural question to ask is,  how can this be extended
to the massive modes? There are two complications - one is that
when massive modes are involved it is hard to put them all
on-shell or even approximately on-shell. Thus one has to learn to deal 
with off-shell fields. The second complication is that of gauge
invariance.  

Some of these questions have been addressed in
 \cite{HLP,KPP,BGKP,Wi,LW}.  In \cite{Wi} an aaproach based
on BRST invariance of the world sheet was developed. In this paper, we do not
have much to say about this approach.  Some of the issues that have
to be addressed in this approach are discussed in \cite{LW,Sh,KMM}.

For the open string, we argue that the loop variable approach gives
an answer to this question. 
At the  free level, equations were written down in \cite{BSLV}. 
A prescription for the interacting case was given in \cite{BSLV0,BSpuri}
and many details were worked out in \cite{BSLV1,BSLV2,BSWF}.

In \cite{BSGI} a simplification was introduced that made gauge
invariance and also other technical issues
 much more transparent. It is easy to show that
the final sytem of equations has the
property of being gauge invariant off shell. The relation between these
equations and the equations that
produce the correct scattering amplitudes for the on-shell physical states,
is the same as that between the Wilson renormalization group equations
with finite cutoff and the Callan-Symanzik $\beta$-function. 
Thus one can expect that when one solves for the irrelevant operators
one will reproduce the on-shell scattering amplitudes. 
This method would thus seem (at tree level) to be an alternative to BRST 
string field theory
\cite{WS,SZ,Wi2}.

This paper explains the results of the loop variable
approach in the form given in \cite{BSGI}. It is a self-contained
discussion. In order to make it self contained it reviews not only
earlier work on loop variables but also contains a short
discussion on the Renormalization Group as discussed by Wilson.
It then attempts to relate the loop variable approach
to the
Renormalization Group (RG) and also other standard concepts from field theory. 
In earlier papers on loop variables, many of the constructs were
ad-hoc and introduced without any justification
other than that they gave the right answer. This paper
attempts to motivate some of these ideas by explaining the connection
with standard field theoretic constructions. Thus, for instance,
the $t$-dependence that was introduced \cite{BSpuri} in the loop momenta
to make the theory interacting has a very transparent
interpretation in terms of  ``sources'' used
in defining generating functionals.
 This paper also attempts to explain in some
detail the connection between the loop variable approach and the usual 
$\beta$-function approach in terms of Wilson's explanation
of the RG \cite{W,WK}. 

This paper is organized as follows. Section 2 discusses some general 
facts about the connection between Wilson's RG and the Callan-Symanzik-
Gell-Mann-Low $\beta$-functions. It also discusses the role of the irrelevant
operators and applies these ideas to the world-sheet RG in string theory.
The c-theorem and some applications to cases that can be
done in a  non-perturbative way (namely, constant e-m field and tachyon
with quadratic profile)
are also discussed here. Section 3 discusses
the same points for the general marginal perturbation, which can 
only be done perturbatively. Section 4 introduces loop variables
as a way of treating all the modes at one go. It also explains the
meaning of some of the constructions, by using loop variables to derive
 some of the results
used in  section 2. Section 5 discusses gauge invariance
and gives the general solution to the problem posed in the introduction.
Section 6 contains some conclusions.

\section{Connection with Wilsonian RG}

\subsection{Generalities about the RG}

The discussion here is based on \cite{WK,W,P,W2}. 
Let us begin by assuming that we have a two dimensional  action, that
corresponds to an open string moving in a completely general open string
 background.
The action has the generic form:

\br    \label{1.1.1}
S~ &=&~ {1\over2}\int _\Gamma d^2\sigma \{\partial ^\alpha X^\mu 
 \partial _\alpha X_\mu \}~+~
\int _{ \partial \Gamma} dt L_1 \nonumber \\    
L_1~&=&~ \sum _i g^i M_i + \sum _i w^i W_{i} + \sum _i \mu ^i R_i
\er 

$\mu$ runs from $0~~ -~~ D-1$. $D$ is 26 for the bosonic string. $d^2 \sigma$ is the area
element in real coordinates and $dt$ the line element. 
Here $\Gamma$ denotes the (Euclidean) world sheet. Thus at tree level
$\Gamma$ is a disc (or upper half plane). $\partial \Gamma$ denotes the boundary
of $\Gamma$. Thus $d^2 \sigma = dxdy$ and $dt=dx$ for the upper half plane.

$L_1$ corresponds to the boundary action corresponding
to  condensation of open string modes. One could also include corrections to the 
bulk action that correspond to closed string modes
in this discussion. Nothing would really be altered. But for concreteness
we restrict ourselves to open string backgrounds, which are boundary terms.
We denote by $M_i ~,~ W_i$, and $ R_i $,  marginal, irrelevant and relevant 
operators
respectively. $ g^i ~, w^i ~, \mu ^i$, are the corresponding coupling constants.
In string theory $g$ would correspond to any mode that is on shell, satisfying
$p^2=m^2$. The off-shell modes would constitute $w,\mu$. At zero momentum,
thus, all the massive modes are irrelevant, the photon is marginal,
and the tachyon is relevant.

The theory is defined with an ultraviolet cutoff, $\Lambda$. Thus the partition 
function
is 
\be
\int _{|p| < \Lambda} [dX(p)]exp \{-S[X(p), g_i,w_i,\mu _i]\} 
\ee

 It is convenient for the
purposes of this section to deal with a finite RG ``blocking'' transformation
that takes the cutoff $\Lambda$ to $\Lambda \over 2$, rather than 
making an infinitesimal change. Denote it by $\cal R$. 
Thus $\cal R$ is to be implemented as follows:\newline
1. Perform the integral
 $\int _{{\Lambda \over 2}<|p|<\Lambda}[dX(p)]exp\{-S[X(p)]\}$.
\newline
2. Rescale momenta: Let $p'= 2p$. Now the range of $p'$ is again $0-\Lambda$.
\newline
3. Rescale the surviving $X(p), ~ 0<|p|<{\Lambda\over2}$. Let
$X(p) = Z X'(p')$. Choose $Z$ so that the kinetic term $p^2 X(p)X(-p)$ has the
same normalization as before.

As a result of all the above we get an expression for the partition function
\be
\int _{|p'| < \Lambda} [dX'(p')]exp \{-S[X'(p'), g'_i,w'_i,\mu ' _i]\} 
\ee

which is exactly the same as before except that the coupling constants have
different values. Thus effectively
\be
{\cal R }~:~ (g,w,\mu )  \longrightarrow (g',w',\mu ')
\ee
defines the discrete renormalization group transformation.

If one plans to iterate the transformation many times one can index
it as

\be    \label{RG}
{\cal R }~:~ (g_l,w_l,\mu _l )  \longrightarrow (g_{l+1},w_{l+1},\mu _{l+1})
\ee
This defines a recursion relation.

A fixed point would be defined by
\be
{\cal R }~:~ (g^*,w^*,\mu ^*  )  \longrightarrow (g ^*,w ^*,\mu ^*)
\ee

In general there are an infinite number of couping constants, labelled by
the superscript $i$ (see (\ref{1.1.1})), but for the purposes of this 
discussion we assume that
$i$ takes just one value.

The recursion relation, then, would for instance take the form
\br   \label{RR}
\mu _{l+1} &=& 4 \mu _l + N_\mu [\mu _l,g_l,w_l] \nonumber \\
g_{l+1}&=& g_l + N_g [\mu _l,g_l,w_l] \nonumber \\
w_{l+1}&=&{1\over 4}w_{l} + N_w [\mu _l,g_l,w_l]
\er

where the factor 4 characterizes a dimension-2 relevant operator
 (eg. a mass term, $X^2$)
 and the factor
1/4 characterizes a dimension-4 irrelevant operator, say of the form
 $(\p X \p X)^2$. $N_\mu , N_g $ and $N_w$ correspond to higher order
corrections and would have to be small if perturbation theory
is to be trusted.

Thus solving for the fixed point would involve setting the coupling
constants to satisfy 

\br     \label{beta}
g_l&=&g_{l+1} \nonumber \\ w_l&=&w_{l+1} \nonumber \\ \mu_l&=&\mu_{l+1}
\er
Thus if the coupling constants are chosen to be their fixed point 
values, then doing
a block transformation does not change anything. This means that 
there are efectively
no dimensionful physical quantities with which to compare the 
cutoff $\Lambda$. Thus
the theory has an overall scale - the cutoff, $\Lambda$, and 
no other scale. 
In particular the correlation
lengths  must be infinite. 

It is important to note that $w^* \ne 0$ in general. Thus it is not true
in general that the massive modes have zero vev's. Why is it that
one usually studies sigma-models with only massless modes turned on?
As explained in \cite{W,WK}, 
one can eliminate all the other modes from
the {\em equations} (\ref{RR})  for the fixed point by solving for them,
 and the resultant
equation involves only the marginal couplings. In order to make this discussion
self-contained we summarize the arguments of \cite{W,WK}:

Let $0\le l \le L$ be the range of the index $l$.  Thus $\mu _0, g_0, w_0$
are the parameters of the action at high energies. From eqn. (\ref{RR})
it is clear that to lowest order $\mu _L \approx 4^L \mu _0$.  This diverges
rapidly with $L$. Thus $\mu _0$ has to be tuned very accurately for the long distance
effective parameter to have some observed value. This is the famous
``fine-tuning'' problem. Thus in solving the equations perturbatively
 we use $\mu _L$ as
our input.  $w_0$, on the other hand keeps getting smaller so it can be used
as an input parameter. This way it can be seen easily that $\mu _l, w_l, g_l$ 
rapidly lose their dependence on $w_0$ and this is a statement of universality.
$g_l$ is important for all values of $l$. 

Now let us iterate the equations (\ref{RR}) a number of times. In the case of 
$\mu _l$ the result looks like
\be
\mu _l = 4^{l-L} \mu _L - \sum _{n=l}^{L-1} 4^{l-(n+1)}N_\mu [\mu _n, g_n, w_n]
\ee 

because we start from $\mu _L$ and go backwards. For $w_l$ we get
\be
w_l=4^{-l} w_0 + \sum _{n=0}^{l-1} 4^{n+1-l}N_w[\mu _n, g_n, w_n] 
\ee

For $g_l$ we start from $g_{l_0}$ and get
\be
g_l=g_{l_0} + \sum _{n=l_0}^{l-1}N_g[\mu _n, g_n, w_n]
\ee

We can solve this equation iteratively with the following starting inputs
obtained by neglecting the non-linear corrections:
\br
\mu _l &=& 4^{l-L}\mu _L \nonumber \\
w_l &=& 4^{-l} w_0 \nonumber \\
g_l &=& g_{l_0}
\er

If one solves this set in terms of the input parameters $\mu_L,w_0,g_{l_0}$
one expects a solution of the form
\[
g_l= V_g(g_{l_0},\mu _L,w_0,l,l_0,L)
\]
and similarly for $\mu _l,w_l$.

However if we are in a region where $l>>0$ and also $l<<L$
(which of course means $L>>0$) then the equations and solution
simplify. Namely the dependence on $\mu_L$ and $w_0$
of $g_l$ is so weak ($O(4^{l-L}$ and $4^{-l}$) that we can
set $\mu_L = w_0=0$ with negligible error. Furthermore
the summations can be extended to $+\infty$ for
$\mu _l$ and $-\infty$ for $w_l$.  The resulting equations
have a translational invariance in $l$ and $l_0$.
Thus 
\be   \label{CSGL}
g_l= V_g[l-l_0,g_{l_0}]
\ee

In physical terms this means that the theory has no dimensionful
parameters in this region: Extending the range of $n$ to $-\infty, +\infty$
means the infrared cutoff is zero and UV cutoff is infinity. The only
other dimensionful parameters, viz $\mu_L$ and $w_0$ have been set to zero.
The resultant solution for $g_l$ can only depend on dimensionless numbers, viz $g_{l_0}$
and the ratio of the two scales
involved, which is $2^{l-l_0}$.  
In this region the recursion relation can be converted to a differential equation, 
 the usual
Gell-Mann - Low, Callan-Symanzik $\beta$-function. The solution of this gives
us $g^*$. One can then solve for $w^*$ if one wants.

In string theory for low energy phenomena
 this calculation thus does not require turning on any 
massive  background fields.
If we are interested in low energy phenomena all massive modes are off-shell
and do not contribute to the $\beta$-function of the marginal couplings.

What happens if we include a massive mode from the beginning? As long
as it scales as an irrelevant operator (which it does at low energies)
 it is really irrelevant. It doesn't affect the 
end point of the flow (which is the fixed point). 
What this means
in practice is that in the usual continuum calculation the operator must have the
appropriate powers of cutoff in the denominator so that it only renormalizes
the marginal coupling and doesn't introduce any new divergences as
the cutoff is taken to infinity.  
However, we repeat: the fixed point value of the irrelevant coupling constant, $w^*$,
 is {\em not} zero,
and the best way to see this is to look at the exact recursion relations (\ref{RR}).
What can be set to zero is the initial value of the irrelevant coupling $w_0$ - as
discussed above. Whether $w^*$ is of the same order as $g^*$ or not
depends on the details of the equations.

In string theory, as an example, consider the massive mode $S^{\mu \nu \rho}$ whose
equation of motion might typically look like \[ (p^2-m^2) S^{\mu \nu \rho} 
\approx A^\mu A^\nu A^\rho .\] Let us work with dimensionless field variables.
Thus we assume that the sigma-model is written in terms of a dimensionless coordinate
$X'^\mu$ where,
$X^\mu = \sqrt {\al '} X'^\mu$. To get fields with canonical
dimensions we multiply the field by appropriate powers of
$\al '$.  Thus for instance in four dimensions $A'^\mu = \sqrt {\al '} A^\mu$
and $S'^{\mu \nu \rho} = \sqrt{\al '} S^{\mu \nu \rho}$ are dimensionless fields.  
The sigma-model
couplings in terms of dimensionless variables are
are $\int A^{'\mu}\p _z X^{'\mu}$ and $\int S^{'\mu \nu \rho } \p _z X^{'\mu}
\p _z X^{'\nu} \p _z X^{'\rho}$.  The equation of motion then is 
\be
\al ' (p^2-m^2) S^{'\mu\nu\rho} \approx A^{'\mu} A^{'\nu} A^{'\rho}
\ee 

If $A^\mu$ has ordinary (weak interaction scale) values then
$A^{'\mu} \approx \eps$ is very small and also
$\al ' p^2 \approx O(\eps ^2)$. But $\al' m^2 \approx 1$.
Let us denote $\al' (p^2-m^2) = \delta$. Then $S' \approx O({\eps ^3\over \delta})$.
So if $\delta \approx O(1)$ then $S' \approx O(\eps ^3)<< A'$
 On  other hand if $\delta \approx O(\eps ^2) $ then 
$S' \approx O(\eps )\approx A'$.
Similarly if $A' \approx O(1)$ then $S' \approx O(1)$.
Thus the moral of this discussion is that the 
massive modes can be important for large values of
$A$ or for high energies but otherwise they are not significant
numerically.

At the fixed point, the physics does not depend on $\Lambda$ so it
can be taken to infinity - the continuum limit. In any approximate
expression
however one can expect explicit powers of the cutoff
and it is best to leave $\Lambda$ finite. 
As an example consider the tachyon vertex operator
$\int dx e^{ikX} \Lambda$. The power of $\Lambda$ indicates
that it is a relevant operator. If one includes various powers of
$\Lambda$ coming from contractions, one gets
$:e^{ikX}: [1+ {k^2\over 2} ln ~\Lambda + ....]\Lambda$.
The expressions $\Lambda ln ~\Lambda$ does not have a well defined
limit as $\Lambda \to \infty$,
for any value of $k$. However if one sums the entire series, one gets
$\Lambda ^{{k^2\over 2} +1} :e^{ikX}:$. For $k^2=-2$, it becomes 
independent of $\Lambda$. Thus the physics becomes independent of the
cutoff at the point $k^2=-2$, on summing all the loops. But to any finite 
order one should
keep $\Lambda$ finite. This is also the message of the $\epsilon $-expansion
\cite{W2}. This also illustrates why, in off shell string theory
(i.e. when $k^2 \ne m^2$), it is important to keep $\Lambda$ finite.
This is because different terms in the interacting equation come with
different powers of $\Lambda$. It is difficult to make sense
of such an equation except when $\Lambda$ is finite.
However the series may sum up to some closed form
expression in which it may be possible to take $\Lambda $ to $\infty$.
Certainly if one has an exact solution to the interacting equation
one should find  that $\Lambda$-dependence disappears.

Thus the conclusion of this discussion is that while writing down the
exact RG equation, one needs to keep a finite cutoff and furthermore
the equation will involve all the massive modes. The fixed point
will have, in general, non-zero values for the irrelevant couplings,
which are the massive modes. In the exact expressions for physical
correlators, evaluated at the fixed point, the cutoff can be taken to
 infinity.
But in any power series expansion the cutoff must be kept finite.
The role of a finite cutoff in string theory RG was illustrated
for the tachyon equation in \cite{BSPT}.

\subsection{Connection with String Theory}

As discussed above, on-shell string fields correspond to marginal
coupling constants for the sigma model and off-shell ones, either
to relevant or irrelevant ones. At zero momentum, all the massive 
fields correspond to irrelevant couplings. The tachyonic fields are
relevant couplings.  
From the discussion of the previous discussion it is thus clear that
on-shell string fields  behave very differently from off-shell fields.

Sigma model methods have typically been used for marginal fields. This
could be near on-shell tachyons, or low momentum vectors.
Consider the following action [$M_i$ are marginal operators]:

\be      \label{1.2}
S~ =~ {1\over2}\int _\Gamma d^2\sigma \{\partial ^\alpha X^\mu 
 \partial _\alpha X_\mu \}~+~
\int _{ \partial \Gamma} dt  \sum _i g^i M_i(t)
\ee

 We know that the following relation is true:
\be    \label{Zamo}
{\p S \over \p g^i} = \beta ^j G_{ij}
\ee
 Here $\beta ^i = {\p g^i \over \p ln~\Lambda }$ is the $\beta$-function for 
the coupling constant $g^i$ and $G_{ij}$ is the Zamolodchikov metric:
\be     \label{ZM}
\lan M_i (z) M_j(w)\ran = {G_{ij}\over (z-w)^2}
\ee
 
$S$ is an ``action'' for the couplings. So in string theory this would be the 
space-time action for the fields of the string. The above is the
 open string version of the corresponding relation for a general 2-D conformal
field theory,
which was first shown in \cite{Zam}.
In \cite{BSPT} it was shown to be true for the open string tachyon, 
to all orders
in perturbation theory. This proof followed closely the outline
of a proof presented in \cite{Poly} for closed strings fields in general.
The proof can presumably be easily generalized to other operators
as well.  An outline of a general proof for this is also given in \cite{KMM}.

Using this relation one can proceed as folows. Let $Z[g^i]$
be the partition function corresponding to the action (\ref{1.2}).
\be
Z[g^i]= \int _{|p|<\Lambda} [dX(p)]exp \{-S[X(p),g^i]\}
\ee

Then 
\be    \label{1.14}
\Lambda {\p \over \p \Lambda} {\p \over \p g^i} Z[g^i]=
\beta ^j {\pp \over \p g^j \p g^i} Z[g^i]=
\int dz \int dw {1\over (z-w)^2} \beta ^j G_{ij}
\ee

Thus for marginal operators we get precisely the equation of motion multiplied by
 an overall factor, the integrals over $(z-w)^{-2}$.
More generally, the operator $\Lambda {\p \over \p \Lambda}$,
produces something proportional to the $\beta$ function, and thus
setting $\Lambda {\p \over \p \Lambda} Z= \int dw \beta ^i \lan M_i (w)\ran =0$
 ensures that we
are at a fixed point, but it does not produce the full equation of motion.

For off-shell fields we need some generalization. $\lan O_i(z) O_j(w)\ran$
for general (not necessarily marginal) operators $O_i$,
 does not satisfy (\ref{ZM}). One needs a generalization
of the Zamolodchikov metric. As an example one can consider
\be    \label{GZM}
\tilde G _{ij} = {1\over L^2}\int _0 ^L dz \int _0 ^L dw (z-w)^2 \lan O_i(z) O_j(w)\ran
\ee 

Clearly for (on-shell) exactly marginal fields $\tilde G_{ij}=G_{ij}$. Something similar was 
considered
in \cite{Wi,LW,Sh,KMM} There the calculation was done on a disc and the 
prefactor was
$sin^2 {(\theta _i - \theta _j)\over 2}= |z_i-z_j|^2$, where 
$z_i=e^{i\theta _i}, z_j = e^{i\theta _j}$
 are points
on the disc. This prescription was derived \cite{Wi,LW} in a very
 elegant manner
using BRST invariance. 
However clearly (\ref{1.14}) will not work. One needs to be able to insert
a factor of $(z-w)^2$ before doing the integral over $z$. 
In the loop variable approach we will use a different
prescription that, as of now, has only been used in a perturbative expansion.
This perturbative method is described in Section 3.  

We will illustrate some of these ideas with two examples that are 
exactly computable.
 One involves
marginal operators. This is the case of the uniform electromagnetic field. 
The other
involves relevant operators and is the tachyon with a quadratic profile.

\subsubsection{Born-Infeld}

The ingredients for this calculation are to be found in \cite{FT,ACNY}.
 $g^iO_i = \int dk A_\mu (k) \p _z X^\mu e^{ik.X}$ is added to the boundary.
In the limit that $A_\mu (k)$ represents a uniform electric/magnetic field,
this problem can be done exactly. Thus
\[
\beta ^\nu = (I-F^2)^{-1\lambda \mu}\p _\lambda F_{\nu \mu}
\] 
is the $\beta$-function to lowest order in derivatives.
Similarly 
\[
\lan \p _z X^\mu (z) \p _w X^\nu (w) \ran = {1\over (z-w)^2} \sqrt {Det(I+F)} 
(I-F^2)^{-1\mu \nu} \equiv {G_{\mu \nu }\over (z-w)^2}
\]
is the Zamolodchikow metric \cite{BSZ}. 

The product gives 
\be     \label{BI}
\sqrt{Det(I+F)}(1-F^2)^{-1\sigma \nu} (1-F^2)^{-1\lambda \mu} \p _{\lambda}
F_{\nu \mu} = {\delta S\over \delta A_\sigma}
\ee

where $S= \sqrt {Det(I+F)}$ is the Born-Infeld action.

The derivation of (\ref{BI}) will be done later using loop variable techniques.
Here we show how the leading terms arise. 
This will illustrate the general arguments.

If $Z[A] = \lan 1 \ran _A$ is the partition function in the presence of the
boundary term, we have
\br
{\p \over \p A_\mu} Z[A]&=& \lan \p _x X^\mu e^{ik.X} \ran _A  \\
&=& \lan \p _z X^\mu e^{ik.X} \{ 1+ \int dk'\int dw ~\p _w X^\nu e^{ik'.X} 
A_\nu (k') + ...\} \ran \nonumber
\er

The first non zero term is thus
\[
\int dw \int dk'~\lan \p _z X^\mu e^{ik.X}   \p _w X^\nu e^{ik.X}A_\nu (k') \ran
\]

We now have to operate with the RG operator $\Lambda {\p \over \p \Lambda}$. 
The $\Lambda$ dependence comes from coincident two point function
which is given by ($a={1\over \Lambda}$),
\be      \label{XX}
\lim _{z\to w}\lan X(z) X(w)\ran = \lim _{z\to w}{1\over \pi} ln~(z-w)
= {1\over \pi }ln ~a =  -{1\over \pi }ln ~\Lambda
\ee

This comes from self contractions in a vertex operator. If we assume
that the vector is transverse, then these are of the form
\[
\int dw \int dk' \lan {k^2\over 2} {1\over \pi} ln~ \Lambda
 :\p _z X^\mu e^{ik.X}: 
\p _w X^\nu e^{ik'.X}A_\nu (k') \ran +
\]
\[
\int dw \int dk' \lan  :\p _z X^\mu e^{ik.X}: 
{k'^2\over 2} {1\over \pi} ln~ \Lambda \p _w X^\nu e^{ik'.X}A_\nu (k') \ran
\]
\be
= (k^2+k'^2) {ln ~\Lambda \over 2\pi} \int dw \int ~dk' \delta(k+k')
 {\delta ^{\mu\nu}A_\nu\over (z-w)^2}
\ee

The RG operator thus gives
\be     \label{17}
\alpha '~2~k^2 A_\nu (k) =0
\ee

We have used the convention $2\alpha'\pi =1$
This is of course the leading term in the equation of motion for a
 transverse vector.

We now go to the next order term:
\[
\int dw \int du \int dv \lan  \int dk'  \p _z X^\mu e^{ik.X}~
 {F^{\alpha \beta}\over 2} X^\alpha \p _u X^\beta (u) ~
{F^{\gamma \delta}\over 2} X^\gamma \p _v X^\delta (v) ~
  \p _w X^\nu e^{ik'.X}A_\nu (k') \ran
\]

We have brought down three powers of the operator $O_i$, and taken the 
uniform field limit for two of them. By keeping a fixed ordering 
$z < u < v < w$, we can
forget about the factor of $1\over 3!$.

Consider the case where the $z,u,v$ operators are first contracted
 amongst each other.
 We get $X^\rho X^\sigma \over 2$
 from  $e^{ik.X(z)}$. Since every contraction can be done in two ways,
 we get a combinatoric factor
of 8 that cancels the 8 in the denominator. 
This gives 
\[
\int dw \int du \int dv
\p _z X^\mu k^\rho k^\sigma \delta ^{\rho \beta} \p _u G(u-z) 
\delta^{\alpha \delta} \p _v G(v-u) \delta ^{\gamma \sigma} G(v-z) 
\]
\[
 F^{\alpha \beta}F^{\gamma \delta}
\p _w X^\nu A_\nu (k') \delta (k+k'){1\over (z-w)^2}
\]

Here $G(z-w) = {1\over \pi}ln~(z-w)$ and satisfies 
$\int du \p_u G(u-z)\p _vG(v-u)= \delta (v-z)$.
Thus we get $\int dv G(v-z) \delta (v-z)= G(0) = {1\over \pi} ln~a = 
-{1\over \pi }ln ~\Lambda$.
Acting with the RG operator gives
$k^\sigma F^{\sigma \delta}F^{\delta \rho} k^\rho \delta ^{\mu \nu} 
\int dk' A_\nu (k') dw {1\over (z-w)^2}$.
One gets a similar contribution replacing $z \leftrightarrow  w$.

Thus the net effect is to replace $k^2$ in (\ref{17}) by
 $k^\rho (I+F^2) ^{\rho \sigma }k^\sigma$
which is the expansion of $k^\rho {(I-F^2)}^{-1\rho \sigma }k^\sigma$. 
The factor of $\sqrt {Det(I+F)}$ comes from the disconnected vacuum to vacuum
amplitude, and multiplies the above expression.
In this manner one can build up the full equation of motion.

\subsubsection{Tachyon}

We now consider an off-shell field: A tachyon with profile
$\Phi (X)= {T_0\over 2\pi} + {1\over 2}uX^2$. This was worked out in 
\cite{Wi,LW,Sh,KMM}.
The world sheet used was a disk.

Using
(\ref{XX}) we see that $\beta _{T_0}= -(u+T_0)$ and
$\beta _u = -u$. Thus to obtain equations of motion we
need the generalizations of $G_{uu},G_{T_0T_0}, G_{u,T_0}$.
We can use for instance the generalized Zamolodchikov
metric  (\ref{GZM}). The vertex operators are $O_{T_0}=1$ and $O_u=X^2$.

We will do the calculations on the Upper Half Plane. For a conformally
invariant theory this should give the same results as on a disk \cite{Wi,LW}.
 However
when one is away from the fixed point, i.e. off-shell, one can expect
differences from \cite{Wi,LW}. However, for large $u$, one can expect that
the fact that the correlation of $X$ dies out rapidly at long distances 
would imply
insensitivity to the shape of the world sheet. Indeed the partition function
calculated on the UHP matches with that on the disk as $u \to \infty$.

 In calculating 
\[
G_{T_0T_0} = {1\over L^2} \int _0^L dz \int _0^L dw (z-w)^2 \lan 1~~1\ran
\]

 and 
\[
G_{uT_0}= {1\over L^2} \int _0^L dz \int _0^L dw (z-w)^2 \lan 1~~X^2(w)\ran
\]

since there is no $z,w$ dependence in the correlation function,
the insertion of $(z-w)^2$ just gives an overall factor of $L^2\over 6$
compared to the expression without the factor of $(z-w)^2$. The expressions
without this insertion are given by ${\pp Z\over \p _{T_0}^2}$ and
$ {\pp Z \over \p _{T_0} \p _u}$.

\[
G_{uu} = {1\over L^2} \int _0^L dz \int _0^L dw (z-w)^2 \lan X^2 (z) ~~X^2(w)\ran
\]
is affected because the connected part of the correlation function involves
$z-w$. 

The leading order term (in powers of $a$ where $a $ is an UV regulator
defined in Sec 4.1)
 in the partition function $Z[u]$, on the UHP is given by (neglecting $T_0$):
\be      \label{PF}
ln ~ Z[u] ~=~ R\int _0^u du' e^{u'a }E_i (-u'a ) + b
= {R\over a}[-ln~ (ua ) -C] + Re^{ua} E_i (-ua ) + b
\ee
$b$ is a constant of integration.  This is derived in Sec 4.1. The original derivation
in \cite{Wi} was done on the disc. In the limit $u\to \infty$ the results are the same.

In the limit $a \to 0$ we get
\be
ln~ Z[u]~=~R u[ln~(ua ) +C] -Ru + b \equiv Ru(F-1) + b \equiv W+b
\ee

 Putting back the $T_0$ dependence we get
\be
Z[u] = e^{-T_0} e^{W+ b}
\ee
\[
{\pp Z\over \p _{T_0}^2}~=~ Z ; {\pp Z \over \p _{T_0} \p _u}= -FZ;
 {\pp Z \over \p _u^2}~=~ Z({1\over u} + F^2)
\]
$Z\over u$ is the connected part.

Thus $G_{T_0T_0} = {L^2\over 6} {\pp Z\over \p _{T_0}^2}= {L^2\over 6} Z$ and
 $ G_{T_0u} = {L^2\over 6}{\pp Z \over \p _{T_0} \p _u} = -{L^2\over 6} FZ$.
The connected part of $G_{uu}$,  becomes $\approx {1\over  u^3}Z$ rather than
 $1\over u$ in $ {\pp Z \over \p _u^2}$.  Thus in the limit
of large $u$,  $uG_{uu} = uZF^2+ {ZF^2\over u^2}\approx uZF^2$.

The equations of motion (in the large $u$ limit) are thus:
\br
{\p S \over \p u}&=&-(F^2-F)uZ + T_0 FZ  =0 \nonumber \\
{\p S \over \p T_0} &=& (F-1)uZ -T_0Z =0
\er

If we plug this in the expression for $Z$, we find that $Z=e^b$.
By appealing to the small $u$ limit one can fix the normalization of
$Z$ (which fixes $b$), as in \cite{Wi,LW}.
These equations are also the same as obtained in \cite{Wi,LW}. 
One can also see that the two equations are mutually consistent.

Thus we have seen the renormalization group and some of its generalizations
being applied in two exactly soluble cases. In the next section we will
discuss the general situation where things are not exactly soluble.
This will lead us to a different off-shell prescription. 
One can also try and use the same prescription as \cite{Wi,LW,Sh} for the general
case. However it is not obvious how to retain BRST invariance in the presence
of a finite cutoff, which is required  in the intermediate
stages of a calculation of the equations of motion (but which presumably is not
required if the fields satisfy those equations).

\section{Equations of Motion for Marginal Perturbations}

\subsection{The General Case}

The Veneziano amplitude for scattering of N particles has the following form:
\be   \label{V1}
\int  ~[dz_3.....dz_{N-1}]\big[ 
\lan V(z_1) V(z_2) V(z_3)........V(z_{N-1}) V(z_N)\ran
(z_1-z_2)(z_2-z_N)(z_1-z_N)\big]
\ee
 If we set $z_N=0$ and take $z_1\to \infty $ this becomes 
\be    \label{V2}
z_1^2 z_2 \int  ~[dz_3 ...dz_{N-1}]\big[  
\lan V(z_1) V(z_2) V(z_3)........V(z_{N-1}) V(0)\ran \big]
\ee
Let us compare this with the following:
\be     \label{PT}
z_1^2 \int dz_2\int ~[dz_3 ...dz_{N-1}]
\big[  
\lan V(z_1) V(z_2) V(z_3)........V(z_{N-1}) V(0)\ran \big]
\ee

If one does the usual rescaling $z'_i = {z_i\over z_2}$ As $z_2 \to 0$, we get a log
divergence 
of the form $ln ~ (z_1/a) $ where $a \approx {1\over \Lambda}$ is some
 lattice spacing 
cutoff. The coefficient of this 
log divergence is the Veneziano amplitude (\ref{V2}). This is understood 
as follows:
 On shell, the dependence of the correlator
on $z_2$ is $1\over z_2$.  In (\ref{V2}), this is cancelled by the factor
 of $z_2$ in front. 
In (\ref{PT}), instead of multiplying by $z_2$ we integrate the amplitude
 over $z_2$ to get 
$\int _a^{z_1}  {dz_2 \over z_2} = ln~({z_1\over a})$. If we extract the coefficient
of this log divergence we get the original amplitude.  Similarly, there is
 a dependence
of the form $(z_1-z_2)^{-1}$, which produces a log divergence as $z_2 \to z_1$, 
with the same coefficient.
All other integrals are also regularized,
 and it can be shown
that this has the effect of subtracting out all intermediate poles \cite{BSPT}. 
This is the proper time method of extracting the equations of motion for marginal
operators (on-shell fields) developed in \cite{BSPT}.

If we considered situations where the operators are not precisely marginal
one gets $z_2^{-1+\delta}$, with $\delta <<1$ instead of $z_1^{-1}$. In this case
we get on doing the integral:${z_1^\delta - a^\delta \over \delta}$. The operation
$a{\p \over \p a}$ brings a factor of $\delta$, and in the limit of small $\delta$,
we have $a^\delta \approx 1$. Typically in these calculations $\delta = p^2-m^2$.
However note that for $N=2$ it gives the kinetic term, because there is no integration over
$z_2$. At higher orders $N\ge 3$, the factors of $p^2-m^2$ cancel
 between numerator and denominator, as discussed 
above, and we are left
with the corresponding Veneziano amplitude - minus some of the poles. The poles
due to the on-shell particles 
get subtracted by the regularization of the integrals. Along these lines, in \cite{BSPT}
it was shown that one gets the correct equation of motion for the tachyon to all orders
in perturbation theory, and also that this equation is proportional to the
 $\beta$-function, with the
proportionality being the Zamolodchikov metric.

\subsection{Tachyon}

Let us illustrate the above ideas with the tachyon \cite{DS,BSPT}.
For a tachyon, $V(k_i,z_i)=e^{ik_i.X(z_i)}$. The perturbation that is
added to the sigma-model  action is $\int _{\p \Gamma} dt~\int dk \phi (k) e^{ik.X(t)}$.

\subsubsection{2-point Function}

We consider 
\br
& &{\p \over \p ~ln~ (a)} \int d k_2 \lan {e^{ik_1.X(z_1)}\over a}
 {e^{ik_2.X(0)}\over a} \ran \phi (k_2) \nonumber \\
&=&{\p \over \p ~ln~ (a)} \int d k_2 e^{{(k_1+k_2)^2 \over 2}ln ~a}
{e^{k_1.k_2 {1\over \pi}ln ~(z_1/a)}\over a^2} \phi (k_2) \delta (k_1+k_2)\nonumber \\
&=&{\p \over \p ~ln~ (a)} e^{({k_1^2\over \pi} - 2) ln ~a} \phi (k_1) e^{-{\ki ^2\over \pi}ln~ z_1} \nonumber \\
\Rightarrow ({k_1^2\over \pi} - 2) \phi (k_1) &=& 2(\alpha ' k_1^2 -1) \phi (k_1)=0
\er

We have introduced the $a$-dependence due to self-contractions
 and also used momentum conservation.
Since we are using the normalization  
$2\pi \alpha ' =1$, the equation above says 
that the tachyon mass is ${1\over \alpha '}$.

\subsubsection{3-point Function}
This gives a quadratic term in the equation of motion.

We consider
\br
& &{\p \over \p ~ln~ (a)} \int ~d k_2 \int ~dk_3 \int _a ^{z_1-a}~dz_2 ~\lambda
\lan {e^{ik_1.X(z_1)}\over a} {e^{ik_2.X(z_2)}\over a}
 {e^{ik_3.X(0)}\over a}\ran \phi (k_2) \phi (k_3) \nonumber \\
&=&{\p \over \p ~ln~ (a)} {1\over a^2} \int d k_2 \int dk_3 \nonumber
\er
\br
& &
\int _a ^{z_1-a}~{dz_2\over a}~
({z_1-z_2\over a})^{k_1.k_2} ({z_2\over a})^{k_3.k_2} ({z_1\over a})^{k_1.k_3}
 a^{{(k_1+k_2+k_3)^2\over 2}}\delta (k_1+k_2+k_3) \phi (k_2) \phi (k_3) \nonumber\\
&=& 
{\p \over \p ~ln~ (a)} \int d k_2 ~
a^{-3 + k_1.k_3 + k_2.k_3 + k_2.k_3} \nonumber
\\ & &
[B(k_1.k_2 +1,k_2.k_3+1)- B_{a/z_1} (k_1.k_2+1,k_2.k_3+1)-
B_{a/z_1}(k_2.k_3+1,k_1.k_2+1)] \nonumber \\ & & \times \phi(k_2)\phi (-k_1-k_2)\nonumber \\
&=&2 \lambda \int ~d k_2~ \phi (k_2)\phi (-k_1-k_2)
\er 

$B_x(a,b)$ is the incomplete Beta-function.
\be     \label{IB}
B_x(a,b) = \int _0 ^x t^{a-1}(1-t)^{b-1} dt = {x^a\over a} 
\big[1+{a(1-b)\over a+1} x+...\big]
\ee

It is understood that we are taking the limit where the particles are all on shell.
In this limit $-3 + k_1.k_3 + k_2.k_3 + k_2.k_1 \to 0$, 
and there is no $a$-dependence. However
when multiplied by 
the Beta function which has poles, a finite number, namely 2, 
results for the coefficient of $ln ~a$. 
This implies a $\lambda {\phi ^3 \over 3}$ interaction for the tachyon.

\subsubsection{4-point Function}
We consider

\br
& &{\p \over \p ~ln~ (a)} \int ~d k_2 \int ~dk_3 \int ~dk_4
\int _a ^{z_1-a}~dz_2 \int _a^{z_2-a} dz_3 \nonumber
\\& & ~\lambda ^2
\lan {e^{ik_1.X(z_1)}\over a} {e^{ik_2.X(z_2)}\over a} {e^{ik_3.X(z_3)}\over a}
{e^{ik_4.X(0)}\over a}\ran \phi (k_2) \phi (k_3) \phi (k_4) \nonumber
\\
&=&{\p \over \p ~ ln~ (a)} {1\over a^2} \int ~d k_2 \int ~dk_3 \int ~dk_4
\int _a ^{z_1-a}~{dz_2\over a} \int _a^{z_2-a} {dz_3 \over a} \nonumber
\\ & &  
[({z_1-z_2\over a})^{k_1.k_2} ({z_1-z_3\over a})^{k_1.k_3} ({z_2-z_3\over a})^{k_3.k_2} 
({z_1\over a})^{k_1.k_4} ({z_2\over a})^{k_2.k_4}({z_3\over a})^{k_3.k_4} ] \nonumber
\\ & & \times
\delta (k_1+k_2+k_3+k_4) \phi (k_2) \phi (k_3) \phi (k_4) \nonumber
\\
&=&{\p \over \p ~ln~ (a)}  {1\over a^2} \int ~d k_2 \int ~dk_3 \int ~dk_4
({z_1\over a})^{ k_1.k_3+k_2.k_3 + k_2.k_3+  k_1.k_4 + k_2.k_4 + k_2.k_4+2} \nonumber
\\& &
 \times \int _{a\over z_1}^{1-{a\over z_1}}dz_2'(1-z_2')^{k_1.k_2} 
{z'}_2^{k_2.k_4+k_2.k_3+k_3.k_4+1}a^{(k_1+k_2+k_3+k_4)^2} \nonumber
\\& &
\times [B(k_3.k_4+1, k_2.k_3+1) - B_{a/z_2}(k_3.k_4+1,k_2.k_3+1) -
B_{a/z_2}(k_2.k_3 +1,k_3.k_4+1)] \nonumber \\ & &\times \phi (k_2) \phi (k_3) \phi (k_4)
\er

For convenience we have assumed a configuration where $k_1.k_3 \approx 0$. 
This factorizes the integral into two parts.

The exponent of $a$  vanishes on shell. Once again the pole terms in the $z_2$ 
integral compensate and we get a factor of 2
mutiplied by:
\be
\lambda ^2 \int ~ dk_3 dk_4 \big[ B(k_3.k_4+1, k_2.k_3+1)- {1\over k_3.k_4+1} - 
{1\over k_2.k_3+1}\big] \phi (-k_1-k_3-k_4)\phi (k_3)\phi (k_4).
\ee

This is the usual Veneziano four-tachyon amplitude minus the on-shell pole parts.

These calculations illustrate the general idea expressed in Section 3.1.
\subsection{Vector}

If one restricts oneself to the physical states (transverse) of the vector
the calculation simplifies and becomes very similar to that above. We illustrate 
with the example of the three point fiunction:
\br
& &{\p \over \p ~ln~ (a)} {1\over a^2} \int _a ^{z_1-a}~dz_2 \int ~ dk_2 \int ~ dk_3 
\\& &
\lan 
\p _{z_1} X^\mu (z_1) e^{ik_1.X(z_1)} \p _{z_2} X^\nu (z_2) e^{ik_2.X(z_2)}
\p _{z_3} X^\rho (0) e^{ik_1.X(0)}  
\ran A^\nu (k_2) A^\rho (k_3) \nonumber
\\
&=&
{\p \over \p ~ln~ (a)} {1\over a^2} ({z_1\over a})^{k_1.k_3} \int _a ^{z_1-a}~{dz_2 \over a}
 ({z_1-z_2\over a})^{k_1.k_2-1} ({z_2\over a})^{k_2.k_3-2} \nonumber
\\ & &
\times k_2^\mu A(k_2).A(k_3)\delta (k_1+k_2 +k_3)a^{(k_1+k_2+k_3)^2\over 2} \nonumber
\\ 
&=&{\p \over \p ~ln~ (a)} {1\over a^2}\int ~ dk_2 \int ~ dk_3 ({z_1\over a})^{k_1.k_3+k_1.k_2+k_2.k_3 -2}
k_2^\mu A(k_2).A(k_3)\delta (k_1+k_2+k_3) \nonumber
\\ & &
\times
 [B(k_1.k_2,k_2.k_3-1) - B_{a/z_1}(k_1.k_2,k_2.k_3-1)
 - B_{a/z_1}(k_2.k_3-1,k_1.k_2)] \nonumber
\er

We have assumed transversality of the vector. Furthermore we have picked a particular
Wick-contraction that corresponds to a particular term in the equation of motion.

We assume that $k_1^2=k_3^2=0$ and $k_2^2 =\eps$ to evaluate the expression before taking $\eps \to 0$.
We use the expansion (\ref{IB}).
 Since we know that the final answer has a log divergence on shell, we keep the coefficient
of the log divergence only and throw away everything else and find the same answer, 2.
This thus corresponds to the $[A,A][p,A]$ coupling in Yang-Mills theory, for transverse states. 
It vanishes in electrodynamics
for symmetry reasons.
If one assumes gauge invariance, one can recover the full coupling from this.

Both, the tachyon and vector calculation described in this section dealt with
on-shell physical states. If the procedure is gauge invariant then the rest of
 the terms in the equation of motion are guaranteed to be correct. The loop variable
method (of Sec 5) is one such gauge invariant procedure and gives a gauge 
covariantized version of the equations derived using the methods of this section.
 Thus on-shell it is guaranteed
to give the right answer.
The important
thing is that it is gauge invariant off-shell also. 
Thus it can be used for all the modes at the same time without losing
gauge invariance.  

\subsection{Connection with c-theorem}

Before concluding this section we make a connection between the proper-time
equation of this section and the method given in the last
section using Zamolodchikov's c-theorem.

In \cite{BSPT} instead of extracting the $ln~a$ dependence, the $ln~ z_1$ dependence was
 extracted. Since $z = e^\tau$, this is the ``proper time''
formalism for strings. It is also shown there that, to all orders, the equation can 
be written as a
 product of the $\beta$ - function and the Zamolodchikov meric, 
as required by the $c$-theorem. One can see the connection as follows:

If $O_i$ and $O_j$ are almost marginal operators then one has
\be
\lan O_i (z) O_j(0)\ran = {G_{ij}\over z^2}
~+~ {H_{ij}\over z^2} ~ln~({z\over a}) + O \big[ ln^2~ ({z\over a}) \big]
\ee

where $G_{ij}$ is the Zamolodchikov metric and $H_{ij}$ are the logarithmic deviations
from scaling that were calculated in the previous section. If we multiply both
sides by $\phi ^j$ and sum over $j$ we get
\be
\lan O_i (z) O_j(0)\ran \phi ^j~=~ {G_{ij}\over z^2}\phi ^j
~+~ {H_{ij}\over z^2} \phi ^j ~ln~({z\over a}) + O \big[ ln^2~ ({z\over a}) \big]
\ee

The coefficient of the log in the second term is precisely the proper time
equation studied in the last section and is also equal to
$G_{ij}\beta ^j$.  Thus we get
\be
\lan O_i (z) O_j(0)\ran \phi ^j~=~ {G_{ij}\over z^2}\phi ^j
~+~ G_{ij} \beta ^j ~ln~({z\over a}) + O \big[ ln^2~ ({z\over a}) \big]
\ee

Since $\beta ^j = -{d \phi ^j \over d ~ln~a}$ we see that in the second
term $\beta ^j ~ln~ (z/a) \approx \delta \phi ^j$.
Thus the equation is essentially
\be
\lan O_i (z) O_j(0)\ran \phi ^j~=~ {G_{ij}\over z^2}(\phi ^j + \delta \phi ^j )
\ee

Thus we can interpret this equation as follows. If we assume that $z$ is close
to $a$ so that we can neglect the higher order terms, then the evolution in $z$
of the state can be seen as a renormalization group evolution of the 
coupling constants of the corresponding sigma-model.

It is not possible to calculate the unintegrated two-point function from the
partition function. It is therefore convenient to calculate the generating
functional, which is nothing but the partition function in the presence of
a very general background. This leads us to the loop variable of the next section.

\section{Loop Variables}

The loop variable method is designed to deal with all the modes at one go.  However
many of the concepts are best illumined by studying some exactly
soluble examples, viz. the tachyon with a quadratic profile \cite{Wi,LW} and also the
constant electromagnetic field \cite{FT,ACNY}.

\subsection{Tachyon}

We illustrate some of the basic ideas using a simplified example of the tachyon.
The best way to deal with very general backgrounds is to use the
Fourier transform. Thus define
\[
\phi (X) = \int ~dk~ \phi (k) e^{ik.X}
\]
If we solve the problem using $e^{ik.X}$ as the background,
then any other background can be treated simply by a suitable
integration over the Fourier transform field. 
 A loop variable is an infinite collection of such ``vertex operators''.
Let us illustrate some of the manipulations using this ``vertex operator''
by applying it to the situation considered in \cite{Wi,LW} and also in the last section,
 namely a tachyon with
a quadratic profile. We will redo the calculation in the upper half plane 
 as an illustration, thereby also deriving (\ref{PF}).
The main difference between this and the usual vertex operators
is that $k$ is not a constant but a current density or a source  (See eqn. (\ref{k})
below).

\be      \label{tach}
S~ =~ {1\over2}\int _\Gamma d^2\sigma \{\partial ^\alpha X^\mu 
 \partial _\alpha X_\mu \}~-~
\int _{ \partial \Gamma} dt  {1\over 2}uX_B(t)^2
\ee 

Here $X_B(t)$ is the value of $X(x,y)$ at the boundary. For the UHP
we can write $X_B(t)= X(x,0)$.
Defining,

\be
S_0~ =~ {1\over2}\int _\Gamma d^2\sigma \{\partial ^\alpha X^\mu 
 \partial _\alpha X_\mu \}
\ee

we get the partition function correponding to the original action (\ref{tach})
as follows:
\be      \label{k}
Z[u] ~=~ \int ~{\cal D}k~ \int ~{\cal D}X ~exp ^{ -S_0[X(x,y)] ~+~ i\int _{\p \Gamma}
~dt ~k(t)X_B(t) }~ \Psi [k(t),u]
\ee
where we have defined a wave-functional,
\be    \label{Psi}
\Psi [k(t),u]= Det ^{-1/2}[2\pi u]~ e^{-{1\over 2u} \int _{\p \Gamma} ~dt~ k(t)^2}.
\ee
We can thus write $Z[u]$ in an obvious way as:
\be    \label{Z}
Z[u] = \int ~ {\cal D}k~ W[k(t)] ~\Psi [k(t),u]
\ee

$W[k(t)]$ is a generating functional. In the terminology used in this paper
$e^{i\int _{\p \Gamma} k(t) X(t) dt}$ is the interacting loop variable
for the case where there is only one mode - the tachyon. 

Thus, once we compute $W[k(t)]$, we can calculate the partition function for any background
simply by multiplying by a wave functional appropriate for that background and
integrating over $k$. Thus all information about any specific background is contained
in $\Psi$. $W[k]$ needs to be computed only once. Let us do that now.

We write $W$ as
\be         \label{W}
W[k]~=~ \int ~{\cal D}X_B \underbrace {\{ \int ~{\cal D}X ~e^{-S_0}~
 \delta [X(u,0) - X_B(u)]\}}_{F[X_B]}
 ~e^{i\int ~du~ k(u) X_B(u)}
\ee 
We are using $u,v ~:~v \ge 0$ as the coordinates of the upper half plane.

In order to evaluate $W$, we write $X(u,v) = X_c (u,v) + x(u,v)$ where
$X_c$ is a solution of the equations of motion satisfying the boundary condition.
Thus let $G(z,w)$ be the Green's function that satisfies
\[
\p  \bar \p G(z,w) = \delta ^2 (z-w)
\]
and \[
G(z,w)|_{y=0}=0\]  We are using the complex notation $z=x+iy,w=u+iv$. 
In real notation
\[
\p _\alpha \p ^\alpha G(x,y;u,v)= 2\delta (x-u)\delta (y-v)
\]
The solution is 
\[
G (z,w) ~=~ {1\over \pi} (ln~ |z-w| - ln~|z-\bar w|)
\]

Then
\be
X_c(x,y) = {1\over 2} \int ~du~ \p _vG(x,y;u,v)|_{v=0}~ X_B(u)
\ee

We can write (after integrating by parts and using properties of the
Greens function):
\br   \label{F}
F[X_B]~&=&~ e^{{1\over 2}\int _{\p \Gamma} ~dt~ X_c \p _n X_c }  Det ^{-1/2} [\pp ]\nonumber \\
&=&e^{ -{1\over 2\pi} \int ~dx~\int ~du~ X_B(x) {1\over (x-u)^2}X_B(u)} Det ^{-1/2} [\pp ]
\er

If we let 
\be   \label{K}
K(x,u) ~=~ {1\over \pi} ln~ (x-u)
\ee , then 
\be   \label{K-}
K^{-1} (x,u) = {1\over \pi }{1\over (x-u)^2}
\ee 
$K$ is, in fact, equal to  $G_N (x,0;u,0)$ where $G_N(x,y:u,v)$ is the 
Green function satisfying
$\p _\al \p ^\al G(x,y;u,v) = \delta (x-u)\delta (y-v)$,  and the
 Neumann boundary conditions
on the real axis :$\p _y G(x,y)|_{y=0}=0$.

Thus we get
\br    \label{W1}   
W[k] &=& \int ~{\cal D}X_B~ F[X_B] ~e^{i\int ~dx~ k(x) X_B(x)}   \\
&=& e^{-{1\over 2} \int ~dx ~\int ~du~ k(x) K(x,u) k(u) } Det ^{1/2} [K]
\er

Thus, using (\ref{Z}) and (\ref{Psi}),  we can calculate the partition function and other
quantities derived from it. 
\br
Z[u] ~&=&~\int ~{\cal D}k~ e^{ -{1\over 2} \int ~dx ~\int ~du~ k(x) ~(K(x,u)
 +{1\over u}\delta (x-u))~k(u)} Det ^{1/2} [K] Det^{-1/2}[2\pi u]\nonumber \\
&=&~ Det ^{-1/2}[K+1/u] ~Det ^{1/2}[K] ~Det ^{-1/2}[u] \nonumber \\
&=&~Det ^{-1/2}[K^{-1}+u]
\er
where we have dropped $u$-independent constants. In momentum space $K^{-1}(p) = |p|$
and so we get
\be
Z[u] ~=~ e^{L\int ~{dp\over 2\pi}~ ln~(|p|+u)}.
\ee
Here $L$ is the size of the box and is an infrared regulator. If we use a UV regulator
of the form $e^{-a|p|}$ in $K^{-1}$ we get eqn (\ref{PF}) as the leading approximation.
\be      
ln ~ Z[u] ~=~ R\int _0^u du' e^{u'a }E_i (-u'a ) + b
= {R\over a}[-ln~ (ua ) -C] + Re^{ua} E_i (-ua ) + b
\ee

$E_i (-ua)$ is defined by the equation \cite{GR}:
\br
\int _0^\infty dk {e^{-ka}\over k+u} &=& -e^{ua} E_i (-ua) \nonumber \\
&=& -e^{ua} (C + ln~(ua) + \sum _{n=1}^\infty {(-ua)^n\over n ~n!})
\er

$C$ is Euler's constant.
The above example illustrates some of the basic ideas in using loop variables.

\subsection{Including All Modes}
 Loop variables
are useful when one has not just one background field as in the above example but an infinite
number of them. Thus one is working with the full string field $\Phi [X(s)]$.
Thus we have
\be    \label{phi}
\Phi [X(s)] = \int ~[dk(s)]~ e^{i\int _c ~ds~k(s)X(s)} \Phi [k(s)] 
\ee

But we have to specify how the proper time (or world-sheet time) enters. So
we assume the geometry shown in Fig.1.
\begin{figure}[htbp]
\begin{center}
\epsfig{file=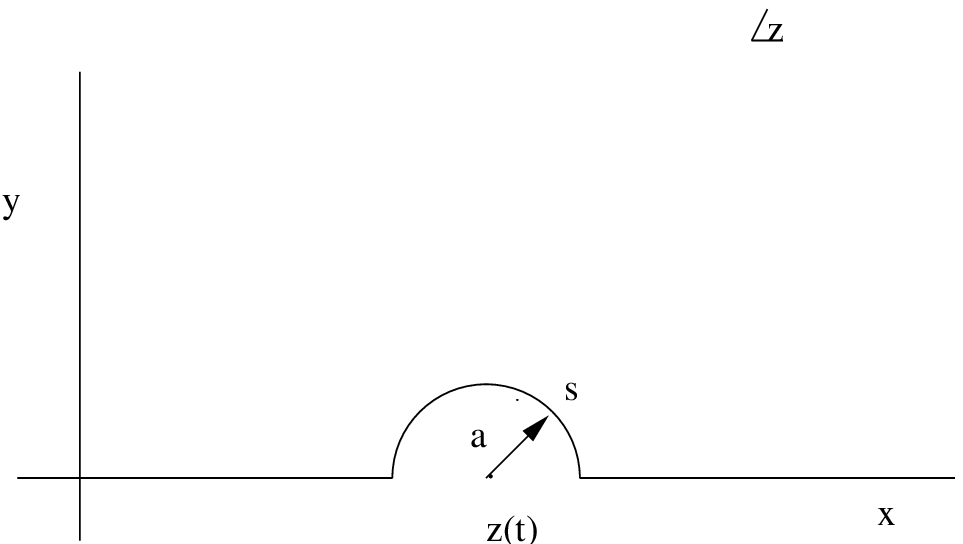, width= 8 cm,angle=0}
\vspace{ .2 in }
\begin{caption}
{The loop variable is a product over all $t$, of variables  defined on a semi-circle of radius $a$ 
centred around a point $z(t)$ on the edge of the upper half plane.}
\end{caption}
\end{center}
\label{Figure 1}
\end{figure}

 The location of the vertex operators 
on the boundary of the world sheet
is $z(t)$. At each point we cut out and remove a small semi-circle of radius $a$. 
The boundary of that semi-circle is thus given by $z~+~ae^{i\theta}=z~+~as$.
Thus we have $X(z(t) ~+~as)$, where $z(t)$ is a point 
on the boundary of the upper half plane, i.e. the real axis. $t$ is a coordinate
that parametrizes the real axis. We can in fact choose $z(t)=t$ \cite{BSWF}.
 \footnote{In some earlier 
papers \cite{BSpuri,BSLV1,BSLV2}
$t$ was used as a continuous {\em index} labelling the various vertex operators
in a correlation function. By setting $z(t)=t$ we are saying that the location itself
can be a label and no additional labels are necessary.}
The generalized momenta
$k(s)$ (or $\kn , n\ge 0$)
 also become a generalized current density $k(s,t)$.
In the limit $a\to 0$ we have a collection of vertex operators, 
all at the same point $z(t)$.   
We have the following Taylor expansion:
\[
X(z+as) ~=~ X(z)~+~as \p _z X(z) ~+~ {1\over 2!} a^2s^2~\pp _z X(z) ~+~....
\]

We also assume that $k(s)$ can be expanded in a power seies in $1/s$.
Thus 
\[
k(s,t) ~=~ k_0(t) ~+~ {\ki (t) \over s} ~+~ {\kt (t)\over s^2}~+~ ....
\]

We use the following definition of the loop variable:
\be     \label{L}
e^{\int _{\p \Gamma} ~dt~i \ko (t) X(z(t)) + 
i \int _{\p \Gamma}~dt~ \int _c ~ds ~k(s,t) \p _z X(z(t)+as)}
\ee

$\p \Gamma$ is the real axis when $\Gamma$ is the upper half plane. We will
suppress this in  subsequent equations. 
In eqn. (\ref{L}) we have separated out the zero mode. 
In the tachyon case we had only the first term.
On performing a Taylor expansion one gets an infinite number of terms from the second term.
The loop variable can be written as

\be
\gvks
\ee

where \[ \tY _n(t) \equiv {1\over (n-1)!}\p _z ^n X(z(t)),~n>0;~~ \tY _0 (t)\equiv X(z(t))
\]

Thus we can define a generating functional
 $W[k(s,t)]~=~W[\ko (t) ,\ki (t) ,\kt (t) ,...\kn (t) ,...]$ (compare (\ref{W}))
by
\br   \label{Wn}
W[\kn (t)] &~=&~ \lan \bp \ran  \nonumber \\
&~=&~ \int ~{\cal D}X(u,v) ~ e^{-S_0[X(u,v)]} \bp \nonumber \\
&~=& \int ~{\cal D}X(u,v) ~ e^{-S_0[X(u,v)]} \gvks
\er

Note that $X(z(t))$ and its derivatives are defined on the boundary, whereas
the integration is over $X(u,v)$ defined on the upper half plane. Thus $X(z(t))$
is actually $X(z(t),0)$.  We can write
(after setting $z(t)=t$) $X_B(t)$ for $X(z(t))$ and  this can be made
 explicit by inserting a delta-function and an 
additional integral over $X_B(t)$ as in (\ref{W}). Information about a specific
background can be encoded in a wave-functional, analogous to (\ref{Psi}), and the
partition function can be computed as a function of all the coupling constants,
which are nothing but the space-time fields of the open string:
\br   \label{Psin}
Z[\phi ^i ]&~=&~ \int ~{\cal D}k(s,t)~ W [k(s,t)] \Psi [k(s,t); \phi ^i]  \\
&~=& \int ~ \prod _{n \ge 0}[{\cal D}\kn (t) ] W[ \ko (t) ,\ki (t) , ..\kn (t), ..] 
\Psi [\ko (t) ,\ki (t),...\kn (t),...; \phi ^i]. \nonumber
\er

In order to avoid confusion we have made explicit all the $t$-dependences
and also the integrations.

\subsection{Uniform Electromagnetic Field}

In  Section 4.1 we saw an example of the wave functional in the case where the
background was a tachyon with a quadratic profile. Another case that is exactly
soluble is the constant electromagnetic field background \cite{FT,ACNY}, and
will be discussed now as another example. This is discussed in \cite{BSWF}.
For simplicity we consider the two dimensional case where we are
given a field strength $F^{01}~= ~F$.
The partition function can then be written as:

\br Z[F]&~=&~  \label{1} 
\int \int {\cal D}\kom (t){\cal D}\kim (t) \nonumber \\&  &
e^{{1\over 2}\int \int dtdt'
[\sum _{n,m=0,1}\km ^\mu (t) {\cal G}_{m,n}^{\mu \nu}(t,t')\kn ^\nu(t')]}~
\Psi [\ko (t),\ki (t), F ^{01}] \nonumber \\
&~=& \int \int {\cal D}\kom (t){\cal D}\kim (t)~ W[\kn ] ~\Psi [\ko (t),\ki (t), F ^{01}] 
\er

Thus we have expressions for 
both the generating functional $W$ and the partition function $Z$.

Here ${\cal G}$ is the matrix
\be    \label{G}
\left(
\begin{array}{cccc}
 K & 0 & \p _{t'}K & 0 \\
0&K&0&\p _{t'}K\\
\p _tK&0&\p _t \p _{t'}K & 0\\
0&\p _t K&0&\p _t \p _{t'}K 
\end{array} \right)
\ee
 where the row vector, $k_m ^\mu$, ($\mu$ is the Lorentz index and $m$ is the
 mode index)  
 multiplying the matrix, is written 
 in the 
following order:
 $(\ko ^0 \, \ko ^1 \,
\ki ^0 \, \ki ^1)$.  $K(t,t') =\langle X(t) X(t')\rangle$ where
$t,t'$ are points on the boundary of the string world sheet. 
 This could be a
disk or the upper half plane. 
$K$  
was calculated in Section 4.1 to be ${1\over \pi} ln~ (t-t')$
 for the upper half plane.
$K$ satisfies the identity 
$\int dt'' \p _tK(t,t'')\p _{t''} K(t'',t')=1$ \cite{ACNY,FT,BSZ}.   

The wave functional in this case is \footnote{The issue
of regularization in defining the wave-functional
is discussed in \cite{BSWF}. Since it does not play a significant
role in this calculation we do not discuss it here.}

\be  \label{1.03}
\Psi [\ko (t),\ki (t);F]=
\int {\cal D}X(t) e^{-i\int dt [\kom (t)X^\mu (t)]} 
\prod _{\mu t}
\delta [\kim (t) - A^\mu (X(t)]
\ee

where $A^\mu = 1/2F^{\mu \nu}X^\nu$, where $\mu =0,1$.

Thus the delta function becomes
\[
\prod _t \delta [X^1(t)-{2\ki ^0(t)\over F}]
\delta [X^0(t)+ {2\ko ^1(t)\over F}]{1\over 
F^2}
\]

Doing the the integral gives
\[
\Psi [\ko (t),\ki (t)]=
e^{{2i\over F}\int dt [\ko ^0(t)\ki ^1(t) - \ko ^1(t)\ki ^0(t)]}
\prod _t{1\over F^2}
\]

When we substitute this expression in (\ref{1}), we get

\be \label{ZF}
\int {\cal D}\ko (t){\cal D}\ki (t)
e^{{1\over 2}\int \int dtdt'
[\sum _{n,m=0,1}\km ^\mu (t) {\cal G}_{F,m,n}^{\mu \nu}(t,t')\kn ^\nu(t')]}
\ee
where 
${\cal G}_F$ is the matrix
\be    \label{GF}
\left(
\begin{array}{cccc}
 K & 0 & \p _{t'}K & {2i\over F} \\
0&K&-{2i\over F}&\p _{t'}K\\
\p _tK&-{2i\over F}&\p _t \p _{t'}K & 0\\
{2i\over F}&\p _t K&0&\p _t \p _{t'}K 
\end{array} \right)
\ee

The Gaussian integral can easily be done and using the identity obeyed by $K$
we get 
\[
Det^{-1/2} [F^{-2}(1+F^{-2})\delta (t-t')]=
\]
\[
 Det[F^2(1+F^2)^{-1/2}\delta (t-t')]
\]

Combining the $1\over F^2$ in the wave functional we get
\[
Det [1+F^2]^{-1/2}\delta (t-t')
\]
Using zeta function regularization (as explained in \cite{FT}) the 
determinant gives
$[1+F^2]^{+1/2}$.
Thus the final answer is 
\be   \label{ZX}
Z[F]= 
[1+F^2]^{+1/2}.
\ee

In order to obtain the equation of motion we can follow the 
prescription of (\ref{1.14}) reproduced here:

\be    \label{14}
\Lambda {\p \over \p \Lambda} {\p \over \p g^i} Z[g^i]=
\beta ^j {\pp \over \p g^j \p g^i} Z[g^i]=
\int dz \int dw {1\over (z-w)^2} \beta ^j G_{ij}
\ee

We thus need to calculate
\br
\Lambda {\p \over \p \Lambda} \lan O_i \ran 
&=&~ \beta ^j \int ~dt'~ \lan O_j (t') O_i(t)\ran \nonumber \\
&=&~ \beta ^j \int ~dt'~ {1\over (t-t')^2} G_{ji}
\er

The operator $O_i$ in our case is $\dot X ^\mu$. We can thus calculate

\br \lan \dot X^\mu (t'')\ran |_A&~=&~  \label{2} 
\int \int {\cal D}\kom (t){\cal D}\kim (t)  \\&~ &
{\delta \over \delta \kim (t'')}~[e^{{1\over 2}\int \int dtdt'
[\sum _{n,m=0,1}\km ^\mu (t) {\cal G}_{m,n}^{\mu \nu}(t,t')\kn ^\nu(t')]}]~
 \nonumber \\
&~&\times \Psi [\ko (t),\ki (t), F ]\nonumber \\ &~=&
\int \int {\cal D}\kom (t){\cal D}\kim (t)~{\delta \over \delta \kim (t'')}
W[\kn ] ~\Psi [\ko (t),\ki (t), F ^{01}] \nonumber
\er

It is quite clear from the symmetry of $W$ and $\Psi$ (both are even in $k$)
that this is zero. Thus the equation of motion is trivial.
When the electromagnetic field is uniform the theory is conformal.
We therefore need to consider next order corrections due to non-uniformity
of the electromagnetic field. We thus take 

\be  \label{1.05}
A^\mu (X) = {1\over 2}F^{\mu \nu}X^\nu + 
{1\over 3} \p _\rho F^{\mu \nu}X^\rho X^\nu
\ee

The delta function in (\ref{1.03}) changes to
\[
\prod _t \delta [\kim (t) - {1\over 2}F^{\mu \nu}X^\nu (t) + 
{1\over 3} \p _\rho F^{\mu \nu}X^\rho (t) X^\nu (t)]
\]
\[
=(1+\int dt'{1\over 3} \p _\rho F^{\mu \nu}X^\rho(t') X^\nu (t')
{\delta \over \delta \ki ^\al (t')} )
\prod _t \delta [\kim (t) - {1\over 2}F^{\mu \nu}X^\nu (t)] 
\] 
 
If we insert this into (\ref{1.03}) we get
for the modified wave functional,

\[
\Psi [\ko (t),\ki (t); F]=
\int {\cal D}X(t) e^{-i\int dt [\kom (t)X^\mu (t)]} 
\]
\[
(1+ \int dt'{1\over 3} \p _\rho F^{\mu \nu}X^\rho(t') X^\nu (t')
{\delta \over \delta \ki ^\al (t')})
\prod _{\mu t}
\delta [\kim (t) - A^\mu (X(t)]
\]

\be   \label{1.07}
=
(1+ \int dt'{1\over 3} \p _\rho F^{\mu \nu}i{\delta \over \delta k_0^\rho (t')}
i{\delta \over \delta \kon (t')}{\delta \over \delta \ki ^\al (t')})
\Psi [\ko (t),\ki (t);F]
\ee

We insert this into (\ref{2}) and integrate by parts the derivatives on $\Psi$,
to get
\[ \lan \dot X ^\al \ran \approx \int ~dt'~\{
[{\cal G} - {\cal G} {\cal G}_F^{-1}{\cal G} ]_{0\nu ,1\mu}(t',t')
[{\cal G} - {\cal G} {\cal G}_F^{-1}{\cal G} ]_{0\rho ,1\al}(t',t'')
+
\]
\[ 
[{\cal G} - {\cal G} {\cal G}_F^{-1}{\cal G} ]_{0\rho ,1\mu}(t',t')
[{\cal G} - {\cal G} {\cal G}_F^{-1}{\cal G} ]_{0\nu ,1\al}(t',t'')
\]
\[
+
[{\cal G} - {\cal G} {\cal G}_F^{-1}{\cal G} ]_{0\nu ,0\rho}(t',t')
[{\cal G} - {\cal G} {\cal G}_F^{-1}{\cal G} ]_{1\al ,1\mu}(t',t'')\} \p _\rho F_{\mu \nu}
\]
\be     \label{calK}
=\{ {\cal K}_{0\nu ,1\mu}(t',t')
{\cal K}_{0\rho ,1\al}(t',t'') +
{\cal K}_{0\rho ,1\mu}(t',t')
{\cal K}_{0\nu ,1\al}(t',t'') +
{\cal K}_{0\nu ,0\rho}(t',t')
{\cal K}_{1\al ,1\mu}(t',t'')\} \p _\rho F_{\mu \nu}
\ee

(This equation defines ${\cal K}$ in an obvious way.)

Using the expressions (\ref{G}), (\ref{GF}) for $\cal G$ and ${\cal G}_F$
one finds:

We are using the following notation: Products of $K$ are understood as
convolution. Thus $KK~ =~ \int dt'~ K(t,t')K(t',t'')$. 
Also  $\p _{t'}K = \p _{t'}K(t,t')$ and $\ddot K = \p _t \p _{t'} K(t,t')$.

\be
[{\cal G} - {\cal G} {\cal G}_F^{-1}{\cal G} ]_{1\mu ,1\nu}(t,t')
=~{\cal K}_{1\mu ,1\nu}(t,t')~=~
\left( 
\begin{array}{cc}
{\ddot K\over 1+F^2} & -i{F \p _{\tau ``} \delta (\tau - \tau ``) \over 1+F^2} \\
i{F \p _{\tau ``} \delta (\tau - \tau ``) \over 1+F^2}  & {\ddot K\over 1+F^2}
\end{array} \right) 
\ee
\be
[{\cal G} - {\cal G} {\cal G}_F^{-1}{\cal G} ]_{0\mu ,1\nu}(t,t')
=~ {\cal K}_{0\mu ,1\nu}(t,t')~=~
\left( 
\begin{array}{cc}
{\p _{\tau '} K\over 1+F^2} & 0 \\
0 & {\p _{\tau '} K\over 1+F^2}
\end{array} \right)
\ee
\be
[{\cal G} - {\cal G} {\cal G}_F^{-1}{\cal G} ]_{0\mu ,0\nu}(t,t')
=~ {\cal K}_{0\mu ,0\nu}(t,t') ~=~
\left( 
\begin{array}{cc}
{ K\over 1+F^2} & -{iFK\p _{\tau '}K\over 1+F^2} \\
{iFK\p _{\tau '}K\over 1+F^2} & {K\over 1+F^2}
\end{array} \right)
\ee

Using these expressions for ${\cal K}$ in (\ref{calK}) one finds
(using $K(t'-t') = ln~ a $, and $\p _t K(t-t') = 0$ when $t\to t'$
by antisymmetry),
\be
\lan \dot X^\al (t'') \ran |_A ~=~ln ~a ~
\int ~ dt'~ \delta _{\rho \nu} \delta _{\mu \al} ({1\over 1+F^2})^2
 {1\over (t'-t'')^2 } \times
\p _\rho F_{\mu \nu}  Z[F] 
\ee

Thus setting ${d\over d ~ln ~ a} \lan \dot X^\al (t'') \ran ~=~0$ 
  gives the Born-Infeld 
equation of motion. Note that
$Z[F]$ is regularized. This is because we need only the term linear in $ln ~a$
and we already have one power of $ln ~a$.

\subsection{General Case: Perturbation Theory}

We have seen how loop variables can be used in two special cases where 
the problem is exactly soluble. For general backgrounds we cannot calculate
$Z[g^i]$ exactly. Nevertheless $W[k(s,t)]~=~W[\kn (t)]$ is always 
the same and can be calculated very easily.

It is in fact given by eqn (\ref{Wn}):
\br   \label{Wn}
W[\kn (t)] &~=&~ \lan \bp \ran  \nonumber \\
&~=&~ \int ~{\cal D}X ~ e^{-S_0[X]} \bp \nonumber \\
&~=& \int ~{\cal D}X ~ e^{-S_0[X]} \gvks
\er

Using the result that $\lan X(t) X(t')\ran ~=~ K(t,t')$ (see (\ref{W1})),
we can write this as
\be   \label{Wnm}
W[\kn (t)] ~=~ e^{-{1\over 2} \int ~dt \int ~dt'~ \sum _{n,m \ge 0} \kn (t) K_{n,m}(t,t') \km (t') }
\ee
where 
\br  \label{Knm}
K_{n,m} &~=&~ \lan \tY _n(t) \tY _m(t') \ran \nonumber \\
&~=&~ { 1\over (n-1)!}{1\over (m-1)!} \p _t^n \p _{t'}^m K(t,t') 
\er

Having defined $W[\kn ]$ we turn to the issue of defining space-time 
fields.
Let us go back to (\ref{phi}) for the moment. This defines (apart from
some minor differences)
the loop variable for a {\em single } string field. In this case $k(s)$ is not
a function of $t$. A Taylor expansion in powers of $s$ yields
all the vertex operators of the open string, but now {\em at one point}
$z$. We consider thus the following definition:

\br
\Phi (X(z+s)) &~=&~ \int ~[d\kn ]~ e^{i\ko .X(z) + \int ~ds~k(s) \p _z X(z+s)} \Phi (k(s))  \\
&~=&~ \int [d\kn ]~e^{i[\ko .X(z) + \ki \p _z X(z) +  \kt \pp X(z) +...+ \kn {\p ^n _z X(z)\over (n-1)!}+...]} \Phi (\kn ) \nonumber \\
&~=&~ \int [d\kn ]~e^{i\ko .X(z)}\{1+  i\ki \p _z X(z) + i\kt \pp X(z) -
\nonumber \\ &~ & {\kim \kin \over 2} \p _z X^\mu (z) \p _z X^\nu (z)+...\} \Phi (\kn )
\nonumber
\er 

In the above expression $\ko$ is the usual momentum of the field. We have all the vertex
operators corresponding to the various modes of the string. This $\kim$ is the analog
of the polarization for $A^\mu(\ko )$. We can define
\br  \label{Free}
\lan \kim \ran &~=&~ A^\mu (\ko ) \nonumber \\
\lan \ktm \ran &~=&~ S_2 ^\mu (\ko ) \nonumber \\
\lan \kim \kin \ran &~=&~ S_{11}^{\mu \nu }(\ko ) 
\er

where $\lan ... \ran ~=~ \int ~d\ki d\kt ..d\kn ..  ...\Phi (\kn )$. 
(Note that $\ko$ is not integrated over.)

$\ko$ is just a number in this definition and one can insert as many powers of $k_0$
 as one likes
on both sides. Thus $\Phi (k(s))$ defines a map from the generalized momenta $k(s)$,
in the loop variable, to space-time fields. 

What is the relation between the string field $\Phi$ defined here and the wave-functional
$\Psi$ defined in earlier sections? For the non-interacting case there is no difference.
Thus when $t$ (and therefore $z(t)$) is fixed $\Psi (\kn ) = \Phi (\kn )$.
However in the interacting case they are different because $\Phi$ always represents {\em one}
string, whereas $\Psi$ stands for an arbitrary number of strings. An example will best
illustrate this. Consider the field $A^\mu (X)$. An open string background 
represents an arbitrary number of fields. In the sigma-model we insert the
following boundary term into the functional integral:
\[
e^{i\int _{\p \Gamma} ~dt~ A_\mu (X(t)) \p _t X^\mu(t)} 
~=~1+{i\int _{\p \Gamma} ~dt~ A_\mu (X(t)) \p _t X^\mu(t)}
\]
\be
 + {1\over 2!}(i\int _{\p \Gamma} ~dt~ A_\mu (X(t)) \p _t X^\mu(t) )^2 
+...
\ee

$A^\mu$ is one mode of $\Phi$. On the RHS therefore we have an arbitrary
number of fields, and therefore an arbitrary number of $\Phi$'s. On the other
hand this entire expression was written in Section 4 in terms of one
wave-functional $\Psi$ in (\ref{1.03}). $\Psi [\kn (t)]$ is a {\em functional}
of $\kn (t)$.  Thus by this device of making all the $\kn$'s a function of $t$,
we can go from the single string Hilbert space
to the multi string Hilbert space. An explicit example of this was given in 
eqn (\ref{1.03}).

Thus in (\ref{Free}) we have to 
replace $\kn $ by $\kn (t)$\cite{BSpuri,BSLV0}.
The loop variable becomes a ``band'' variable.
 Two $\kn$'s with different values of $t$ thus represent
two different strings - an interaction, whereas two $\kn$'s with the same value
of $t$ represent a higher excitation of the same string. This gives us

\br     \label{inter} 
\lan \kim (t) \ran &~=&~ A^\mu (\ko (t))  \\
\lan \ktm (t)\ran &~=&~ S_2 ^\mu (\ko (t) ) \nonumber \\
\lan \kim (t_1)\kin (t_2) \ran &~=&~ S_{11}^{\mu \nu }(\ko (t_1) ) \delta (t_1-t_2)
 + A^\mu (\ko (t_1)) A^\nu (\ko (t_2)) \nonumber
\er

We also have to see the effect of  $\ko (t)$. $\ko (t)$ represents the momentum of
 the string labelled by $t$.

Thus 
\be
\lan \ko ^\nu (t_1) \kim (t_2) \ran ~=~ \delta (t_1-t_2) \ko ^\nu (t_1 )A^\mu (\ko (t_1))
\ee

and similarly

\[
\lan \ko ^\rho (t_1) \kim (t_2) \kin (t_3)\ran ~=~ 
\delta (t_1-t_2) \ko ^\rho (t_2 )A^\mu (\ko (t_1)) A^\nu (\ko (t_3))
\]
\be    
 +
\delta (t_1-t_3) \ko ^\rho (t_3 )A^\mu (\ko (t_1)) A^\nu (\ko (t_3)) +  
\delta (t_1-t_2)  \delta (t_2-t_3)\ko ^\rho (t_2) S_{11}^{\mu \nu }(\ko (t_2) )
\ee

Eqn (\ref{inter}) along with similar equations involving all the other $\kn $'s,
 define the wave functional $\Psi$. For perturbative
calculations we do not need an explicit expression for $\Psi$. We can thus
calculate ${d \over d ln~a} \lan O_i \ran $ as a power series
in the marginal fields in order to get the equations of motion. This
method works for the tachyon and massless vector. 

Note that for the tachyon $O_i= e^{ik.X}$, but due to momentum conservation,
\[
\lan e^{ik.X} \ran = 0
\]
 when $k\neq 0 $. Thus the calculation starts
at second order on bringing down one factor of $\int ~ dz~\int ~dk~ \phi (k) e^{ik.X(Z)}$.
Except for an overall integration over $z$ this is identical to the proper-time
equation described in Section 3. The same is true for the (transverse) vector.
However for massive modes this does not work due to issues of gauge invariance.
One needs a more elaborate loop variable to get gauge invariant results. 
This is the subject
of the next section.

\section{Gauge Invariance}

 \subsection{Free Theory}
In order to obtain gauge invariant equations one needs a local
version of the RG equations where $a$ is not constant
on the world sheet. The following example illustrates this.
One way to obtain the equations is to require that the
cutoff dependence of operators vanish. Thus instead
of evaluating $\lan O_i \ran$, we just evaluate
the cutoff dependence of $O_i$. If one doesn't impose momentum
conservation on the VEV these two calculations are in fact the same.
Thus for the tachyon, ${1\over a}e^{ik.X} = e^{({k^2\over 2\pi }-1) ln ~a }:e^{ik.X}:$. 
Here :..: denotes normal ordering.
Setting ${d\over d ln~a }=0$ gives the equation of motion of the tachyon. Let us attempt
to do this for the vector.
\[
A^\mu (k ) \p _z X^\mu e^{ik .X} = - ik . A(k ) {\p _z \sigma \over 2\pi} :e^{ik .X}: e^{{k^2\over 2\pi} \sigma} + 
A^\mu (k) :\p _z X^\mu e^{ik.X}: e^{{k^2\over 2\pi} \sigma}
\]
where we have used $\sigma$ in place of $ln ~a$ and 
also $\lan \p_zX (z) X(z) \ran = {\p_z \sigma \over 2\pi}$.
 $\sigma$ can be thought of as the Liouville
mode which is a local measure of the scale. We can say that $a \to a e^\sigma$.
Now we vary w.r.t $\sigma$, and  after integrating by parts, set $\sigma=0$ to find,
\[
(-k.A(k) k^\mu + k^2 A^\mu (k)) :\p_z X^\mu e^{ik.X}:
\] 

This is Maxwell's equation. Had we not kept the $\p _z\sigma$
piece we would have obtained only one of the terms and the result would
not have been gauge-invariant. This illustrates the importance of
{\em local} scale invariance.

We can attempt to do the same thing for the massive modes of the form
$S_2 \p _z^2X e^{ik.X}$. In that case
one can expect terms of the form $k.S(k)\p_z^2 \sigma :e^{ik.X}:$. 
On varying w.r.t $\sigma$, when we integrate by parts and
differente  $e^{ik.X}$ twice,
 we get terms of the form
$k.S(k) k^\mu k^\nu :\p _z X^\mu \p _z X^\nu e^{ik.X}:$. This term is
clearly unacceptable in  an equation of motion for a free field
because of the presence of three factors of $k$. The problem gets
worse at higher mass levels. Thus we have to modify our loop variable
\cite{BSLV}.    

The basic idea introduced in \cite{BSLV} is to introduce
additional variables $\xn , n>0$ that, very roughly,
have the property that ${\p \over \p \xn }\approx \p _z ^n$.
This means that when we integrate by parts over $\xn$ (not $z$)
and the derivative acts on $e^{ik.X}$ 
we never bring down more than one power of $k$. 

Another problem with the loop variable is that it depends
on the parametrization of $s$. It is not diffeomorphism invariant.
This can be rectified by introducing 
an einbein along the loop. The modes of this einbein also provide the 
extra variables $\xn$. Thus both problems are solved.

We thus consider the following loop variable:
\be   \label{84}
\lpp
\ee

$\al (s)$ is an einbein. Let us assume the following Laurent expansion:
\be
\al (s) ~=~ 1 ~+~ {\al _1 \over s}~+~{\al _2 \over s^2} ~+~ {\al _3 \over s^3}+...
\ee

Let us define 
\br
Y &~=~& X ~+~ \al _1 \p _z X ~+~ \al _2 \p_z^2 X ~+~
 \al _3 {\p _z ^3 X \over 2}~+~ ...~+~{\al _n \p _z^n X\over (n-1)!}~+~...\nonumber \\
&~=~& X ~+~ \sum _{n>0} \al _n \tY _n \\
Y_1 &~=~& \p _z X ~+~ \al _1 \p_z^2 X ~+~ \al _2 {\p _z ^3 X \over 2}~+~ ...~+~{\al _{n-1} \p _z^n X\over (n-1)!}~+~...\nonumber \\
...& & ...\nonumber \\
Y_m &~=~& {\p _z^m X\over (m-1)!} ~+~ \sum _{n > m}{\al _{n-m} \p _z^n X\over (n-1)!}\\
\er

If we define $\al _0 =1$ then the $>$ signs in the summations above can be replaced by $\ge$.

Using these equations one can write
\be
\lpp = \gvk
\ee

It is understood that $Y_0=Y$.

Let us now introduce $\xn$ by the following:
\be
\al (s) = \sum _{n\ge 0} \al _n s^{-n} = e^{\sum _{m\ge 0} s^{-m} x_m}
\ee

Thus 
\br
\al _1 &=& x_1  \nonumber \\
\al _2 &=& {x_1^2 \over 2} + x_2 \nonumber \\
\al _3 &=& {x_1^3 \over 3!} + x_1x_2 + x_3
\er

They satisfy the property,
\be
{\p \al _n \over \p x_m} = \al _{n-m} , ~~ n\ge m
\ee

Using this we see that 
\be
Y_n = {\p Y\over \p x_n}
\ee

Now we can do the same operation of imposing $\dds =0$.
Except we will define  $\Sigma = \lan Y(z) Y(z) \ran$. This is
equal to the previous $\sigma$ in coordinates where $\al (s) =1$.
Thus we have for the coincident two point functions:
\br    \label{Sig}
\lan Y ~Y\ran &~=&~ \Sigma \nonumber \\
\lan Y_n ~Y\ran &~=&~ {1\over 2}{\p \Sigma \over \p x_n}   \nonumber \\
\lan Y_n ~Y_m \ran &~=&~  {1\over 2}({\pp \Sigma \over \p x_n \p x_m} - {\p \Sigma \over \p x_{n+m}})
\er

Using this the normal ordering gives:
\br   \label{LV}
\lpp &=& \gvk  \nonumber \\
&=& exp \{\ko ^2 \Sigma + \sum _{n >0} \kn .\ko  {\p \Sigma \over \p x_n} +  \nonumber \\
& & \sum _{n,m >0}\kn .\km {1\over 2}({\pp \Sigma \over \p x_n \p x_m} - {\p \Sigma \over \p x_{n+m}})\} \nonumber \\
& & :\gvk :
\er

We can now operate with  ${\delta \over \delta \Sigma}$ and set $\Sigma =0$. We will only give one sample variation here:

\[
{\delta \over \delta \Sigma} [
\kn .\km {1\over 2}({\pp \Sigma \over \p x_n \p x_m} - {\p \Sigma \over \p x_{n+m}})
] :e^{i\ko .Y}:
= :({1\over 2}i\kom i\kon Y_n^\mu Y_m ^\nu + i\kom  Y_{n+m}^\mu ) e^{i\ko .Y}:
\]  

One can thus collect all the coefficients of a particular vertex operator,
say $:Y_n ^\mu e^{i\ko .Y}:$, and this gives the free equation of motion. Note that
they never contain more than two space-time derivatives. This solves the first problem.

The second problem is that of gauge invariance. We have assumed that
$\al (s)$ is being integrated over, which is why we are allowed to integrate
by parts. This means that 
\be   \label{GT}
k(s) \to \la (s) k(s)
\ee
 is equivalent to $\al (s) \to \la (s) \al (s)$, which is clearly just a change
of an integration variable. Assuming the measure is invariant this does nothing
to the integral. The measure ${\cal D}\al (s)$ has been chosen to be $\prod _n d\xn $.
If we set $\la (s) = e^{\sum _m y_m s^{-m}}$ , then the gauge transformation (\ref{GT})
 is just a translation, 
$\xn \to \xn + y_n$ which leaves the measure invariant. Thus we conclude that
(\ref{GT}) gives the gauge transformation. 

If we expand $\la (s)$ in inverse powers of $s$
\[
\la (s) = \sum _n \la _n s^{-n}
\]
Then we can write (\ref{GT}) as 
\be    \label{GT1}
\kn \to \sum _{m=0}^{n} \la _m k_{n-m}
\ee

We set $\la _0 =1$.

In order to interpret these equations in terms of space-time fields 
we use (\ref{Free}). They have to be extended to include $\la$. Thus we
assume that the string wave-functional is also a functional of $\la (s)$. 
Thus we set
\br   \label{La}
\lan \la _1 \ran &~=&~ \Lambda _1 (\ko )\nonumber \\
\lan \la _1 \kim \ran &~=&~ \Lambda _{11}^\mu (\ko ) \nonumber \\
\lan \la _2 \ran &~=&~ \Lambda _2 (\ko )
\er

The gauge transformations (\ref{GT1}) thus become on mapping to
space time fields by evaluating
$\lan .. \ran$:

\br
A^\mu (\ko )~&\to &~ A^\mu (ko ) + \kom \Lambda _1 (\ko ) \nonumber \\
S^\mu _2 (\ko )~&\to &~ S^\mu _2 (ko ) + \kom \Lambda _2 (\ko ) + \Lambda ^\mu _{11} \nonumber \\
S_{11}^{\mu \nu} ~&\to &~ S^{\mu \nu}_{11} + k^{(\mu}_0 \Lambda _{11}^{\nu )}
\er  
These are more or less the canonical gauge transfomations for a massive spin two field.
\footnote{They become identical after we perform a dimensional reduction. This will be described later.}
Now it is known that the gauge transformation parameters of higher spin fields
obey a certain tracelessness condition \cite{F,SH}.  We will see this below also.

When one actually performs the
gauge transformations we find the following mechanism for gauge invariance.
It changes the
normal ordered loop variable by a total derivative in $\xn$ 
which doesn't affect the equation of motion.
More precisely the gauge variation of the loop variable is a term of the form
${d\over d\xn } [A(\Sigma ) B]$, where $B$ doesn't depend
on $\Sigma$. The coefficient of $\delta \Sigma$ is obtained as
\[
\int ~~\delta ({d\over d\xn } [A(\Sigma ) B]) =
\int ~~ ({d\over d\xn}( {\delta A\over \delta \Sigma } \delta \Sigma ) B +
  {\delta A\over \delta \Sigma } \delta \Sigma {dB\over d\xn })
\]
\[
=\int ~~[ - {\delta A\over \delta \Sigma } {dB\over d\xn } +
 {\delta A\over \delta \Sigma } {dB\over d\xn }]\delta \Sigma =0
\]
Here we have used an integration by parts.

Actually one finds on explicit calculation that the variation is not a total
derivative. This is because in deriving (\ref{Sig}) some identities
have been used. Thus only if we use those identities in the variation will
we be able to write the variation as a total derivative. However we do
not want to use them because we would like to leave $\Sigma$ unconstrained
when we vary. Thus constraints have to be imposed elsewhere. It can easily be
checked that the terms that have to be put to zero are all of the form
\be
\la _n \km . k_p ...
\ee   

where ... refers to any other factors of $k_m$ \cite{BSLV,BSLV'}. 
Thus all traces
of gauge parameters have to be set to zero. 
This thus explains the tracelessness
mentioned earlier.

In \cite{BSLV,BSLV'} some examples, namely spin-2 and spin-3
are explicitly worked out.

The form of the gauge transformation (\ref{GT}) is very suggestive:
It is a {\em scale transformation in space-time}. It is local
along the loop. As explained in \cite{BSLV} getting a gauge 
transformation of this type
was one of the original motivations for this formalism.

\subsection{Dimensional Reduction}

In the gauge invariant formalism we have not made contact with 
string theory because we
have not reproduced the mass spectrum.  
In earlier sections we were dealing with the physical
states, and the mass, being the dimension of the operator,
came out right because there are always appropriate
powers of the cutoff $a$ multiplying the operator
 to give it the right dimension. In
the gauge invariant formalism, we are using the new
$\Sigma$ field, which is a complicated combination of the old
$\sigma$ field and its derivatives. Thus we cannot just
introduce it  by replacing $a$ by $ae^\s$, as one would normally do
when introducing the Liouville mode.
We will introduce it by hand as a variant of the Kaluza-Klein
mechanism. Thus we will let the momentum $\kom$ be a 27-dimensional
vector rather than a 26-dimensional one. We will let $k_0^{26}$
stand for the mass as in Kaluza-Klein theories but instead of letting $\ko$ be
multiples of $1\over R$ we will assume that $\ko ^2$ is a multiple
of $1\over R^2$. Furthermore gauge invariance will force us to make all
the $\kn$'s 27-dimensional. This means many new modes are being
introduced into the theory. This is just as well - we know that
the gauge invariant formalism requires  an additional bosonic
coordinate worth of modes \cite{SZ}.

We denote $k^{26}$ by $q$. We set $q_0$ to $\sqrt {(P-1)}$, where
$P$ is the engineering dimension of the vertex operator. Thus
for the tachyon $P=0$, for the vector $P=1$ etc. Finally there is
one subtlety. The first oscilator of the extra bosonic
coordinate was set to zero in \cite{SZ}. In our case
the first mode $q_1$ will not be set to zero identically because
that would violate gauge invariance. We will impose relations
of the form:
\[
\lan q_1 \ran =0 .
\]
\[
\lan q_1 q_1 \ran = \lan q_2 q_0 \ran = S_2 ~; ~\lan \la _1 q_1 \ran = \lan \la _2 q_0 \ran = \Lambda _2 .
\]
\be
\lan q_1 \kim \ran = \lan \ktm q_0 \ran = S_2^\mu .
\ee

Note that $q_0$ is 1 for the spin-2 field. 
There will relations of this type that will enable us to get rid of $q_1$
completely. The form of the relations is such as to maintain gauge invariance.
They can be built up recursively.

We merely summarize the results for the massive spin-2 field:
\[
\delta S_{11}^{\mu \nu} = \ko ^{(\mu } \Lambda ^{\nu )}
\]
\[
\delta S_{2}^\mu  = \Lambda _{11}^\mu + \ko ^{\mu } \Lambda _2
\]
\be
\delta S_2 = 2 \Lambda _2
\ee

These are in the ``standard'' form, where the extra auxiliary fields
$S_2$ and $S_2^\mu$ can be set to zero to recover the Pauli-Fierz
equations for massive spin-2 fields. Further details can be found in
\cite{BSLV} and references therein.

\subsection{Interactions}

We now turn to the all important issue of introducing interactions.
The prescription for interactions that was first proposed in
\cite{BSpuri}, worked out in \cite{BSLV1,BSLV2}, and modified
into a very simple form in \cite{BSGI} is to introduce
an extra parameter $t$ in all the variables. Thus
$k(s) \to k(s,t)$ and $X(z + s) \to X(z(t) +s)$. $t$ is supposed
to label the different strings. This was an ad-hoc procedure then. But
in view of the calculations described in Section 4 where we defined
the partition function in terms of a generating functional
$W[k(s,t)]$ (see eqn (\ref{Wn})), in retrospect the meaning 
of introducing $t$ is very clear.
As the discussion following (\ref{NO}) makes clear we end up calculating $W[k(s,t)]$.
Thus  it is clear that this procedure  gives interactions.
 What we have to ensure is that
gauge transformations can still be consistently defined.

Let us follow the prescription of \cite{BSpuri,BSLV1,BSLV2}.
First of all we need a generalization of $\Sigma$ for the interacting case.
This can be done by considering an alternative definition of $\Sigma$ \cite{BSCD,BSLV1}.

Perform a general conformal transformation, with group element
$\eln$,  on the loop variable:
\be   \label{lv}
\gvk .
\ee

The resulting $\la$ dependence (which is also an expression of the 
$\sigma$ (Liouville mode) dependence), can be rewritten in the form
given in (\ref{LV}).

We use the following result \cite{BSVir}:
\be  \label {Vir}
\eln e^{iK_{m} \tilde{Y}_{m}}
= e^{K_{n}.K_{m}\lambda _{-n-m}+\tilde{Y_{n}}\tilde{Y_{m}}\lambda_{+n+m}
+imK_{n}\tilde{Y_{m}}\lambda_{-n+m}}e^{iK_{m}\tilde{Y_{m}}}
\ee

The anomalous term is $\Kn .\Km \lambda _{-n-m}$ and the classical 
term is $ m\Kn \tilde{\ym} \lambda _{-n+m}$. We will ignore the 
classical piece: this can be rewritten as a $(mass)^{2}$ term, which 
will be reproduced by performing
a dimensional reduction, and other pieces involving derivatives
of $\Sigma$ (defined below), which correspond 
to field redefinitions \cite{BSLV}.  
We can apply (\ref{Vir}) to the loop variable 
(\ref{lv}) by setting 

\be
\Km = \sum _{n} k_{m-n} \aln .
\ee  Defining
\be
\Sigma =  \sum _{p,q} \alpha _{p} \alpha _{q} \lambda _{-p-q}
\ee
we recover (\ref{LV}). This definition of $\Sigma$  
generalizes in a straightforward way  to the interacting case.

 The $\Sigma (z_1,z_2))$
will be defined in terms of $\rho (z_1,z_2)$, which is defined as follows. First
\be     \label{3.15}
e^{:\frac{1}{2}\int du \lambda (u) [\partial _{u} X(u)]^{2}:}
e^{ik_{n}
{\p _{z_1}^n\over (n-1)!} X(z_1)}
e^{ip_{m}{\p _{z_2}^m\over (m-1)!} X(z_2)} 
\ee
defines the action of the Virasoro generators on the two sets of 
vertex operators.
\be     \label{3.16}
= e^{k_{n}.p_{m}{\p _{z_1}^n\over (n-1)!} {\p _{z_2}^m\over (m-1)!}  
\oint du\frac{\lambda (u)}{z_{1}-z_{2}}[\frac{1}{z_{1}-u} -\frac{1}{z_{2}-u}]}
\ee
\[
= e^{\kn .p_m {\p _{z_1}^n\over (n-1)!} {\p _{z_2}^m\over (m-1)!}  \rho (t,t')}
\]
\[
=e^{k_n .p_m \rho _{n,m}(z_1,z_2)}.
\]
where $\rho (z_1,z_2) = {\la (z_1) - \la (z_2)\over z_1-z_2}$.
  $\rho$ is a generalization
of the usual Liouville mode $\s = {d\la \over dz}$, for the case where the
vertex operators are not located at the same point. It has to be further modified
to make it gauge covariant - this will give us $\Sigma$. We will do this later.

In the interacting case when we consider the normal ordering of
 the products of vertex operators, we not only have in the exponent, $\rho$
and its derivatives, but also $K$ (Green's function) and its derivatives.
As a simple illustration consider the conventional normal ordering of the 
product $e^{ik.X(z) + ip.X(w)}$,

\br   \label{NO}
e^{ik.X(z) + ip.X(w)} &=& e^{{1\over 2}\lan (ik.X(z) +i p. X(w)) (ik.X(z) +i p. X(w))\ran}
:e^{ik.X(z) + ip.X(w)}: \nonumber \\
&=& e^{\{ {1\over 2\pi}[(k^2 + p^2 ) ln ~(ae^\s ) +  2k.p ln (z-w)]  }   \nonumber \\
&=& e^{\{ {1\over 2\pi}[(k^2 + p^2 ) ln ~(a ) +  2k.p ln (z-w) + (k^2+p^2) \s ]  }   
\er

In the exponent $\sigma$ is what is generalized to $\rho$. But we also
have $K(z,w)= ln~ (z-w)$. $ln ~a$ is just $K(z,z)$ regularized. A convenient
form for $K$ that can be used is $K(z,w) = {1\over 2\pi}ln~ (a^2 + (z-w)^2)$.

 Notice also that 
if we denote by $V(k,z)$ the vertex operator at $z$ with momentum $k$,
the above equation (\ref{NO}) can be written as
\be  \label{VV} 
V(k,z)V(p,w) = \lan V(k,z)V(p,w)\ran : V(k,z)V(p,w):
\ee
The only subtle point is that there is 
effectively a momentum conserving delta-function in
(\ref{VV}), whereas it is not there in 
(\ref{NO}).  

Thus if our loop variables are of the form
\[
e^{i\int ~ dt~ \sum _{n\ge 0} K_n (t) \tY _n (t)}
\]
we get for the normal ordered expression, the following $\rho ,K$-dependence:
\be
exp \{ \int ~dt~\int ~dt'~ [K_0 (t). K_0 (t') [K(t,t') + \rho (t,t')] + 
\ee
\[
\sum _{n>0}2K_n (t). K_0 (t') [K_{n,0}(t,t')+\rho _{n,0}(t,t')] + 
\sum _{n,m >0} K_n (t).K_m (t') [K_{n,m}(t,t') +\rho_{n,m}(t,t')]]\}
\]
\[
:exp(i\int ~dt~K_n(t) \tY _n(t)):
\] 

As in (\ref{VV}), the coefficient of the normal ordered operator is
\be
\lan e^{i\int ~ dt~ \sum _{n\ge 0} K_n (t) \tY _n (t)} \ran
\ee

This is of course nothing but the generating functional $W[K_n (t)]$! Thus our
prescription of introducing $t$-dependence, makes contact with our
earlier discussions on interacting string theory in Section 4.

Now one has to introduce the $\aln$'s to make the formalism gauge invariant.
At this point the method adopted in \cite{BSpuri, BSLV1,BSLV2} is to introduce
$\aln (t)$ and define a generalized Green's function, $ G(t,t')$,  and generalized Liouville
mode $\Sigma (t,t')$ by taking combinations of the
form $\sum _{n,m}\al (t) _{p-n} \al (t')_{r-m} [K+\rho ]_{n,m}(t,t')= [G+\Sigma ]_{r,p}(t,t')$.
The technical complication involved in this is that $\Sigma (t,t') $ is not a local
field but a bi-local field. A Taylor expansion was then performed. There is also some ambiguity
regarding the $t$-dependence of $\al$.  In \cite{BSLV1,BSLV2} the $t$-dependence was retained
in the intermediate stages of the calculation but in the end the $\al$'s were set
to be $t$-independent.

A simpler alternative was followed in \cite{BSGI} which we will use here. All ambiguities
are avoided by first Taylor expanding all the $X(z)$'s about $X(0)$ in powers of
$z$. Thus all calculations involve the vertex operators located at a single point $z=0$.
\footnote{The choice $z=0$ is not important - it could have been about any point
$z=z_0$ as well.} This is well defined provided we keep a finite cutoff in all intermediate
stages of the calculation so that coincident two-point functions
are finite. Thus we could use for instance $K(t,t') = {1\over 2\pi}ln~(a ^2 + (t-t')^2)$. 
Since the equations involve off-shell vertex operators,
in any case we do need a finite cutoff.

Thus we first write
\be
\sum _{n\ge 0}\kn (t) \tY _n (z(t)) = \sum _{n\ge 0}{\bar \kn (t,-z(t))} \tY _n (0)
\ee

The argument $t$ is superfluous and one can write
$\bar \kn (t,-z(t)) = \bar \kn (-z(t))$. If we let $z(t)=t$ we can further
simplify notation and write just $\bar \kn (-t)$.

This defines $\bar \kn (-z(t))$ to be
\be   \label{2.29}
{\bar k_q (-z(t))} = \sum _{n=0}^{n=q} k_n (t)  D_{n}^{q}(z(t))^{q-n}
\ee
where
\br
D_n^q &  = & ^{q-1}C_{n-1},\; \; n,q\ge 1 \nonumber \\
      &  = & {1\over q}, \; \; n=0 \nonumber \\
      &  = & 1 ,   \;\;           n=q=0
\er

Note that ${\bar \ko }(-z(t)) = \ko (t)$. 

Now we can write the gauge invariant loop variable 
 as
\be   \label{LVinter}
e^{\int ~dt~\sum _{n\ge 0}i{\bar \kn }(-z(t)) \yn (0)}
\ee

One can also rewrite this as a loop variable analogous to (\ref{84}).
Define first, $k(s-z)=\sum _{n\ge 0} \kn (-z) s^{-n}$. Consider
\[
\sum _{n>0}\kn (-z) \tY _n (0)  + \ko X(z) = 
\sum _{n>0}(\kn (-z) + \ko {z^n\over n})\tY _n(0)+ \ko \tY (0)
\]

($\tY \equiv X$.)
(The variable in brackets is in fact $\bar \kn (-z)$ defined earlier in (\ref{2.29}.)
\br
&=&~\sum _{n>0}\bar{\kn} (-z)  \tY _n(0) + \ko (z) X(0)
\nonumber \\ 
&=&~ \int ds ~\sum _{n>0}{\bar \kn }(-z) s^{-n} \p X(s) + \ko (z) X(0)\nonumber \\
&=&~ \int ds {\bar k}(s,-z)\p X(s) + \ko (z) X(0)
\er
This equation defines $\bar k(s,-z)$.
We can now introduce an einbein $\al (s)$ to get
\be   \label{2.213}
\int ds {\bar k}(s,-z)\al (s) \p X(s) + \ko (z)X(0)
\ee

Thus (\ref{LVinter}) (or (\ref{2.213})) is the loop variable we work with. The
interesting thing is that all the $z$-dependence is in $k$
rather than in $X$. Thus we have to work with $Y_n(0)$.  There is
no ambiguity about the $t$-dependence of $\al _n$ since they are
all at one point.  Thus (\ref{LVinter}) looks a lot like (\ref{LV})
except that the coefficients $\kn$ are replaced by
$\int ~dt~ {\bar \kn} (-t)$, and $\Sigma$ replaced by $\Sigma (0,0)+ G (0,0)$.  
This also makes the calculations very similar
to the free case. Only after calculating $W[\kn (t)]$ when we integrate
over $\Psi [\kn (t)]$ the effects of this replacement will show up through
all the $t$-dependences of the correlation functions and subsequent 
integrations over $t$ \cite{BSGI}. This will be shown below.

Thus the normal ordering gives exactly the same result as
(\ref{LV}) with appropriate replacements:

\br   \label{LV'}
e^{\int ~dt~\sum _{n\ge 0}i{\bar {\bar \kn} (-t)} \yn (0)}
&=& exp \{ \int ~dt~ \int ~ dt'~ {\bar \ko } (-t) {\bar \ko } (-t') (\Sigma (0,0) + G (0,0)) + 
\nonumber \\
& &
\sum _{n >0} {\bar \kn} (-t) .{\bar \ko } (-t') 
 {\p (\Sigma (0,0) + G (0,0)) \over \p x_n} +
\nonumber \\
& & \sum _{n,m >0}{\bar \kn} (-t) .{\bar \km } (-t')
 {1\over 2}({\pp (\Sigma (0,0) + G (0,0)) \over \p x_n \p x_m} -
\nonumber \\
& &
 {\p (\Sigma (0,0) + G (0,0)) \over \p x_{n+m}})\} \nonumber \\
& & :e^{\int ~dt~\sum _{n\ge 0}i {\bar \kn} (-t) \yn (0)} :
\nonumber \\
&=& W[\kn (t)] :e^{\int ~dt~\sum _{n\ge 0}i {\bar \kn} (-t) \yn (0)} :
\er

We have already specialized to $z(t)=t$ in the above equation.

The equations of motion can be obtained as in the free case by
operating with $\delta \over \delta \Sigma (0,0)$ and then setting
$\Sigma$ to zero. One can also set $\Sigma=0$ right in the beginning
and perform the operation $\delta \over \delta G(0,0)$ treating
$G$ formally as a field and we get the same answer. Either way
we can see that this is a gauge covariantized version of
$\dds$ (or $d \over d ln ~a$), which is the renormalization group
operator.  This also makes contact with the formalism of Section 2 and Section 3.

We will now define gauge transformations:
\be  \label{GTinter}
{\bar \kn }(-t) \to {\bar \kn }(-t) + 
\int ~dt'~\sum _{m\le n} \la _m (t') {\bar k_{n-m}} (-t)
\ee
This is the same as in the free case except for the replacent
of $\la _n$ by $\int dt \la _n (t)$. Thus the gauge invariance of the
equations follows in exactly the same way as in the free case.
Note that we also have the interacting version of the constraints,
which now have the form:
\[
\int ~dt~\la _p(t) {\bar \kn} (t_1,-z(t_1)).{\bar \km }(t_2,-z(t_2)) ....=0
\]

There will also be equations generalizing (\ref{inter}) to include
$\la _n (t)$ (apply the definitions (\ref{La}). We do not write them out
explicitly.  

\subsection{Defining Gauge Transformations of Space-Time Fields}

Let us say that ${\cal M}$ is a map from an expression $L$ involving
 products of ${\bar \kn} (-t)$ to an expression $S$ involving
products of space-time fields. 

Thus \[
{\cal M} ~:~ L ~~\longrightarrow ~~~S
\]
This is  the operation 
$\lan ... \ran = \int {\cal D}\kn {\cal D}\la _n ...\Psi [\kn (t), \la _n(t)]$.

Let $\cal G$ denote the  
gauge transformation (\ref{GTinter}). Thus 
\[
{\cal G} ~:~ L ~~~\longrightarrow ~~~L^g
\]

Then  ${\cal M}$ maps this to another set of space-time fields -
$S^g$.
\[
{\cal M}{\cal G} : ~L ~=~ S^g
\]

  Let $G$ be the gauge transformation on $S$ induced by this.
Thus 
\be  \label{map}
G~~:~~~ S ~~\longrightarrow ~~{\cal M}{\cal G} ~L~ =~ S^g
\ee

If we define gauge transformations this way we are guaranteed that
any expression $L$ that is invariant under $\cal G$, (i.e. $L_g=0$)
maps to a gauge invariant $S$, (i.e. $S_g=0$). This would mean that
all the loop variable equations that are by construction gauge invariant
will lead to gauge invariant equations of motion. The question is whether 
there is a unique and well defined action of $G$ on {\em individual} fields
such that (\ref{map}) (which involves sums of products of space-time fields)
is satisfied.

We show that the answer is in the affirmative. 
Let us be more precise and specify the steps involved in the argument.
We will fill in the details afterwards.

{\bf Step 1}

Any equation of motion is a sum of terms of the type
\br
 \lan {\bar k}_{n_1}^{\mu _1} (-t_1) {\bar k}_{n_2}^{\mu _2} (-t_2) 
{\bar k}_{n_3}^{\mu _3} (-t_3) ...
{\bar k}_{n_r}^{\mu _r} (-t_r) \ran &=& \lan L_{n_1,n_2,...,n_r} \ran 
\nonumber \\
&=& S_{n_1,n_2,...,n_r}
\er
 where $S_{n_1,n_2,...,n_r}$ is a sum of products of space-time fields and
also various powers of $t_i$.  (All the $t_i$ are integrated at the end
of the day.) Note that the set of numbers uniquely specify $L_{n_1,n_2,...,n_r}$
and vice versa. \footnote{There could be Lorentz-index contractions. This is
trivially taken care of.}

{\bf Step 2}

It is possible to define {\em uniquely} the action of $G$ on space-time fields 
so that the gauge transformation $G$ acting on this expression 
$S_{n_1,n_2,...,n_r}$ is the expression $S_g$ obtained by the action of
${\cal MG}$ on $L_{n_1,n_2,...,n_r}$ as explained above. That is
\[
G ~ S_{n_1,n_2,...,n_r} = S_{n_1,n_2,...,n_r}^g 
\equiv \lan {\cal G} L_{n_1,n_2,...,n_r} \ran
\equiv \lan L_{n_1,n_2,...,n_r}^g \ran
\]

{\bf Step 3}

The loop variable equation of motion is of the form $L_1 + L_2 +...$ by step 1.
If ${\cal G} (L_1 + L_2 +...)= L_1^g + L_2^g + ..=0$ then
$G(S_1 +S_2 +S_3 +..) = S_1^g + S_2^g +...=0$. 
Thus the equations are gauge invariant in terms of space-time fields.

Let us fill in the details:

Step 1:

The expression
 $W[\kn (t)] :e^{\int ~dt~\sum _{n\ge 0}i {\bar \kn} (-t) \yn (0)}
:$ eqn. (\ref{LV'}),  is clearly of the type 
$L_{n_1,n_2,...,n_r}$ (and its Lorentz index contractions.)
When we vary w.r.t. $\Sigma$ we integrate by parts on $\xn$. But
$L_{n_1,n_2,...,n_r}$ does not depend on $\xn$ so it is unaffected.
Thus every equation is made up of terms $L_{n_1,n_2,...,n_r}$.
This ensures that the set of numbers uniquely specifies the combination
that occurs in the equation of motion.
Thus we get also uniquely $S_{n_1,n_2,...,n_r}$.

Step 2:

This is the crucial step.

A given set of numbers $n_1,...n_r$ uniquely defines
the set of terms involved in $S_{n_1,n_2,...,n_r}$. It involves
one highest level field that we can call 
$S^{\mu _1 \mu _2 \mu _3...\mu _r}_{n_1,n_2,n_3..n_r}$ and
products of lower level fields such as
$S_{n_1,n_2}^{\mu _1 \mu _2} S_{n_3,...n_r}^{\mu_3 ...\mu _r}$.
If some of the $n_i$ are zero then we consider an expression $S'$ without
those $n_i$. Then $S$ is obtained from $S'$ by taking space-time derivatives.
This is because the operation $\int dt \ko (t)$ is just differentiation.
So we assume that none of the $n_i$ are zero. 

Given the expression $S_{n_1,n_2,...,n_r}^g$, we can define
recursively the gauge transform of the highest level field
$S^{\mu _1 \mu _2 \mu _3...\mu _r}_{n_1,n_2,n_3..n_r}$,
as $S_{n_1,n_2,...,n_r}^g$ {\em minus} the gauge transform
of the rest of the expression which involves only lower level fields.
This definition is unique (assuming the lower level fields have uniquely
defined transformations). This procedure therefore allows us to recursively
define gauge transforms of higher fields in terms of lower ones
such that $S^g = GS$.

Step 3:

Follows from the definition of $S^g$.
 
--------------

What we really have is an algorithm for defining gauge transformations 
of fields in such a way that the equations of motion are gauge invariant. 

An example illustrating this is given in \cite{BSGI}.
We now turn to the issue of dimensional reduction in the interacting case.

\subsection{Dimensional Reduction in the Interacting Case} 

In the free theory we set $q_0 = \sqrt {P-1}$. In the interacting case
we proceed as follows. Concentrate on the transverse physical state scattering
where all the particles are on shell. We set $q_0 (t_i)$ to the value that
produces the correct scattering amplitude. Then we will show that gauge 
invariance
is not affected by this prescription. Then by gauge invariance 
we are guaranteed that this
will work for the longitudinal and pure gauge states also.  Finally since
gauge invariance is valid off-shell also, we can extrapolate to get
the off-shell equations also. 

To get the prescription for physical states let us consider the
equation of motion for the state $Y_{n_1} ^{\mu _1} Y_{n_2}^{\mu_2}...
Y_{n_r}^{\mu _r}$. The leading term is just 
\[
(p^2+m^2) S^{\mu _1 \mu _2 ...\mu _r}_{n_1 n_2 ..n_r} =0
\]
where $m^2 = n_1 +n_2 + ...+n_r -1$.  Let us set $r=2$ for the moment.
Consider the interaction between $S_{n_1}^{\mu _1}$ and  $S_{n_2}^{\mu _2}$
to produce $S_{n_1,n_2}^{\mu _1,\mu _2}$. 
 so that we have to look at the $a$-dependence in the integrated
correlator of two vertex operators. The number of powers of $a$
from the vertex operators is $(n_1-1)+(n_2-1)$,  the ``classical'' part and
$p^2 +q^2$, the anomalous part. If $z,w$
are the locations, then we have the integrals 
$\int dz  \int dw $. Thus the final answer involves lookng at
the coefficient of $ln ~a$ in
\[
\int {dz \over a} \int {dw \over a}
({z-w\over a})^{2p.q}a^{p^2+q^2+2p.q}   a^{n_1+n_2}
\]
We set all the particles on shell, $p^2 +n_1-1=0$ and $q^2+n_2 -1=0$.
If we also set $(p+q)^2 + n_1+n_2-1=0$ we see that all the powers of
$a$ cancel except those coming from the regularization of the integration
( because $2p.q \approx -1$) which gives a log divergence, 
whose coefficient is 1. 
Alternatively if we keep all momenta slightly off-shell, the integration gives a result
${1\over 2p.q+1}$. When we do the operation $d\over d ~ln ~a$ we get
$(p+q)^2 + n_1+n_2-1$. When $p^2 +n_1=1$ and $q^2+n_2=1$, the numerator
zero cancels the pole and we get 1. 
 This gives the on-shell interaction
between $S_{n_1}^{\mu _1} , S_{n_2}^{\mu _2} , S_{n_1,n_2}^{\mu _1 \mu _2}$.
If instead of including explicit factors of $a^{n_1}$ and $a^{n_2}$ with vertex
operators,
we naively let $p,q$ be 27-dimensional then we will get
$a^{(p^{26}+q^{26})^2}$. This must provide the powers of $a$ that come from the
vertex operators and the final undone integral which is $n_1+n_2-1$. 

Thus we set $(p^{26}+q^{26})^2=n_1+n_2-1$.

The only difference between the above calculation and the loop variable
calculation is the expression $a^{(p+q)^2}$ becomes $e^{(p+q)^2 \Sigma }$ 
in the loop variable
calculation. 

All this is easily generalized to the general case where $r>2$.
We simply set $(\sum _{i=1}^r q_0 (t_i))^2 = \sum _{i=1}^r n_i -1 = P-1$.
Thus our prescription is that in an expression 
$L_{n_1,n_2,...,n_r}$ we must set 
$(\sum _{i=1}^r q_0 (t_i))^2 = \sum _{i=1}^r n_i -1 = P-1$. 
 This can be achieved by setting all the $q_0 (t_i) = \sqrt {P-1\over r}$.

There is one other important point. Notice that in the term
$(z-w)^{2p.q}$ $p,q$ have to be 26-dimensional momenta. So clearly
the 27th component does not modify the $z$-dependence 
of the correlation function.  We will therefore set 
(letting $V$ denote the 26th direction)
$K(z,w) ^{VV}
= \lan Y^{V}(z) Y^{V}(w)\ran =0 ~~~~\forall z\neq w$. This effectively
means setting  $q_n(t,-z(t)) = q_n(t,0)$. This also affects the
$\Sigma $ terms.  We need not worry about the geometrical significance
of what this means for the world sheet conformal theory of the
27th coordinate is concerned.  The point is that gauge invariance is 
not affected because the transformation law (\ref{GTinter}) did
not depend in any way on the $z$-dependence of the $\kn$.
Thus to summarize: the prescription is to set $q_0 = \sqrt {P-1\over r}$
, where $r$ is the number of fields in a term in the equation of motion.
and $q_n (t, -z(t)) = q_n(t,0) ,  ~~\forall n$.

Nor does this affect the consistency and uniqueness arguments
of the previous section for defining gauge transformations of space-time fields.
Provided we include the index $n$ of the gauge transformation parameter
$\la _n$ in calculating $P$. 
This is because the gauge transformations are defined in 
terms of units that we denoted by $L_i$. Thus in a given
$L_i$, $q_0$ has a well defined value which will determine
the gauge transformation. For a different $L_j$, $q_0$ can take
another value. If $L_1^g$ has terms of the same form as
$L_2^g$ then they also necessarily have the same value of $P$ and hence of 
$q_0$
because the level must necessarily be the same. More precisely 
if $L_1^g + L_2^g =0$, this implies that $S_1^g + S_2^g=0$.

Thus at this point we have a set of equations that are correct
when the particles are all on-shell.  Furthermore they are gauge invariant
on-shell and off-shell.  

\subsection{Connection with RG}

We have already seen above that as far as on-shell physical modes
are concerned the equation is obtained
in exactly the same way as would be obtained had we done an RG operation
$d\over d~ln~a$. However to make the formalism gauge invariant off shell
also,  we
had to do a Taylor expansion of $Y(z)$. This means that to get the
$z$-dependence (while keeping gauge invariance) we have to introduce
the higher massive modes. Thus for instance if we want the equation
involving $A^\mu$, which requires $\kim$, we must also use
${\bar k}_2^\mu (-z)$ because this contains in it $z\kim $. Once we
use ${\bar k}_2^\mu (-z)$ we automatically have $\ktm$. Thus massive
modes are forced into the equation when we require gauge invariance
off shell. This has been noticed earlier \cite{BSFC}. 

From the RG point of view this is natural. When we are off-shell
we are away from the fixed point and one expects all the irrelevant
operators to appear in the equation.  These equations are of the form
(\ref{RR}).  The equations of motion for on-shell fields (marginal operators)
is of the form (\ref{CSGL}), that gives the Callan-Symanzik (Gell-Mann - Low).
$\beta$-function.

To see that the equations we get are indeed RG equations in the off-shell case
let us count the powers of $a$ and show that ${\delta \over \delta \Sigma}$ 
counts them. ${\delta \over \delta \Sigma}$ brings down a factor
of $\int dt_1~\int dt_2~(\kom (t_1).\kom (t_2)  + 
q_0(t_1). q_0 (t_2))$, where $\mu :~0-25$. To make things simple
we can work in dimensionless coordinates by replacing $z$ by $z/a$.
The two point function 
$\lan X({z\over a}) X({w\over a}) \ran = {1\over 2\pi}ln ~ (1+{(z-w)^2\over a^2})$.
In the original definition of the loop variable $\yn$ came with a power
of $a^n$. To compensate this we let $\kn \to \kn /a^n$. In this way
it is clear that all the terms in ${\bar k}_n (-z/a) = {\kn \over a^n} +
 {z\over a} {k_{n-1}\over a^{n-1}} +...+ \ko {z^n\over a^n}$ all have the
same dimension $n$. Thus $\int dt_1~\int dt_2~q_0(t_1). q_0 (t_2)$
, following our prescription above counts this dimension $n$. The anomalous
$a$-dependence of $e^{ik.X}$ which is $a^{k^2}$ and the overall $a$-dependence
of $\lan :e^{ik.X(z)}: :e^{ik.X(w)}:\ran = a^{2p.q} (1+ ({z-w\over a})^2)^{2p.q}$    
which is $a^{2p.q}$ is what is counted by $\int dt_1~\int dt_2~\kom (t_1).\kom (t_2)$.

Thus we see that the  term in the equation of motion which comes from
$\int dt_1~\int dt_2~(\kom (t_1).\kom (t_2)  + 
q_0(t_1). q_0 (t_2)) \Sigma $ is the RG equation involving massive as well as
massless modes. The terms coming from derivatives of $\Sigma$ are required
for gauge invariance. Thus we see that we get a gauge invariant
generalization of the
of the exact Wilsonian RG equations by this procedure.

From the above discussion we see that the dimensionful coupling constant
$\kn$ and $\ko z^n$ have the same dimension. Equivalently
the dimensionless coupling can be defined by $\kn '= {\kn \over a^n}$.
This is to be compare with $\ko {z^n\over a^n}$. Now the upper
limit in the range of $z$-integration, which can be denoted by $R$ 
is an infrared cutoff. After doing the integral we have to
compare the irrelevant coupling $\kn$ with the marginal 
$\ko$ which occurs in the same equation as $\ko {R^n\over a^n}$.
Thus in the infrared limit $R\to \infty$ 
(or equivalently, in particle physics terminiology, the continuum limit $a \to 0$)
the contribution of the 
irrelevant coupling is small, as expected. The ratio
$R\over a$ is an expansion parameter and measures the relative
importance of the higher dimension operators corresponding
to the irrelevant coupling constants. This is analogous
to the number 4 in (\ref{RR}) or the level-expansion parameter $4\over 3\sqrt 3$ in BRST
string field theory.

In \cite{BSGI} some of the higher order terms correcting Maxwell's
equations were calculated.

The leading term in Maxwell's equation including the contribution of the tachyon is:

\be   \label{ME}
(a ^2) ^{\ko ^2}[(k^2iA ^\mu (k) - A(k).k ik^\mu ) + (p+r)^2 iA^\mu (p) \phi (r) - 
A(p).(p+r)i(p+r)^\mu \phi (r)]
\ee 

A $z$-independent correction term involving massive modes:

\be   \label{S3}
(a ^2)^{\ko ^2}
(-{4\over a ^2})\{ (-i)\p ^\mu \p ^\rho \p ^\sigma 
[(1+\phi)(S_{2,1}^{\rho \sigma}+ S_2^\rho A^\sigma )] 
+4i\p ^\mu [i\p ^\rho [S_3^\rho (1+\phi)] + 2\sqrt 2 \p ^\mu (S_3 (1+\phi )) 
\ee
We have set $(q_0)^2=2$.

A $z$-dependent term involving massless modes.
\be   \label{zA}
 (a ^2)^{\ko ^2}\{-6{z^2\over a ^2} \mup \pp \p ^\rho [A^\rho (1+\phi )] +2
 {(z -z')^2\over a ^2}\mup \p ^\rho \p ^\sigma [A^\rho \p ^\sigma \phi ]\}
\ee

We first rewrite (\ref{S3}) in terms of dimensionless coupling
$S'_{2,1}= {S_{2,1}\over a^2}$.
Comparing eqn. (\ref{S3}) with eqn. (\ref{zA}), we see that the
contribution of the marginal coupling $A$ is enhanced by the factor $R^2\over a^2$
relative to $S'_{2,1}$.  Also comparing the corrections with the leading order
term we see that an approximate solution will have $\al ' p^2 A(p) \approx
{a^2\over R^2} A(p) \approx S'_{2,1}$.  This shows the relative contributions
of relevant and irrelevant operators.

If we now naively take the limit $a\to 0$ we
will be forced to set the e-m field to be a constant and all massive modes to zero.
This will give us a ``trivial'' fixed point. If on the other hand if we keep 
$a$ finite and include all the higher correction terms with some critical
values for the couplings, there is the possibility that
they all add up to some closed form expression where the cutoff can be taken
to zero. This would give a non-trivial fixed point.

A simple  example of this is very familiar: The higher derivative
terms involving $(z/a)^n$ and $\p ^n A$ (and products of $A$)
are obviously expansions of terms of the form
$(a^2 + z^2)^{p.q}$   that are present in the Veneziano-type amplitude for 
scattering of photons. This is exactly what we get when we perform an operator
product expansion of the the product of two vertex operators (but using the regularized
Green's function). 
The product of two photon operators can be replaced by an infinite sum of massive
mode vertex operators.
 Clearly when the photon is on-shell 
we can undo this expansion - then all these terms add up to some non-trivial but 
finite amplitude with a smooth  limit $a\to 0$. The limit
$a\to 0$ should only be taken after adding all these terms.
 One could also have more non-trivial 
examples such as the well known tachyon condensation \cite{AS2}.

\section{Conclusions}

In this paper we have dealt with the following question: How does
one obtain gauge invariant equations of motion for all the modes
of the open string using the renormalization group? Over and
above the issue of gauge invariance there is the issue of going
off shell. We have proposed a solution to this problem using
loop variables.  A condensed description of this proposal was
given earlier in \cite{BSGI}. We have also tried to make clear
the connection with the exact RG equation of Wilson
and  also the connection with the Callan-Symanzik and Gell-Mann-Low
equations. Zamolodchikov's c-theorem plays a role in
this understanding.
We have illustrated some of the ideas by applying
them to known examples.

Some intriguing features about this
method were also mentioned in \cite{BSGI}. We mention it here also
for completeness.

 First, the theory is formally
written as a massless theory in 27 dimensions and masses are obtained by
a dimensional reduction prescription (that is quite a different one from
the usual Kaluza-Klein reduction). Second, the structure
of the interacting theory, both the form of the equations and the gauge 
transformation law, is similar to that of the free theory. The loop is just 
thickened to a band and the loop variables acquire a dependence on the 
positions of the vertex operators. Third, the gauge transformation law,
in terms of loop variables has a simple interpretaion of space-time scale
transformations. This supports the speculation \cite{BSLV} that the  space-time
Renormalization Group on a lattice with finite spacing, is part of the invariance
group of string theory.
Finally,  space-time gauge invariance of the equations obtained this way
does not seem tied down to any special world sheet properties, unlike in BRST
string field theory where it follows from BRST invariance. To that extant
it need not describe a string theory. Only the special choice of Green's function
enforces the string theory connection. This may be a desirable feature from the 
point of view of the problem of background independence.

We also mention  some of the open problems.
We have not investigated the issue of whether there is a simple generalization
that works for higher order string-loop corrections. It is also not known whether
the equations follow from an action.
The precise relation to BRST string field theory 
is not clear. The theory is so much simpler in terms of loop variables that
it would be interesting to work out solution generating techniques in terms
of these variables rather than in terms of space-time fields.
Finally, it would be interesting to find a physical explanation of the
``intriguing'' features mentioned above.

\end{document}